\documentclass[a4paper,12pt]{article}
\usepackage{amsmath}
\usepackage{amssymb}
\newcommand{\sect}[1]{\setcounter{equation}{0}\section{#1}}
\def\rf#1{(\ref{eq:#1})}
\def\lab#1{\label{eq:#1}}
\def\nn{\nonumber \\}
\newcommand{\beano}{\begin{eqnarray*}}
\newcommand{\enano}{\end{eqnarray*}}
\def\bea{\begin{eqnarray}}
\def\ena{\end{eqnarray}}
\def\foot#1{\footnotemark\footnotetext{#1}}

\relax

\font\fld=msbm10 at 12 pt
\font\goth=eufm9 at 12 pt
\newcommand{\fl}[1]{\mbox{\fld #1}}     
\newcommand{\go}[1]{\mbox{\goth #1}}    

\newcommand\bil[2]{\langle {#1}\>|\> {#2} \rangle} 
\def\ra{\rightarrow}

\def\PRL#1#2#3{{\sl Phys. Rev. Lett.} {\bf#1} (#2) #3}
\def\NPB#1#2#3{{\sl Nucl. Phys.} {\bf B#1} (#2) #3}

\def\CMP#1#2#3{{\sl Commun. Math. Phys.} {\bf #1} (#2) #3}
\def\PRD#1#2#3{{\sl Phys. Rev.} {\bf D#1} (#2) #3}
\def\PRv#1#2#3{{\sl Phys. Rev.} {\bf #1} (#2) #3}

\def\PLB#1#2#3{{\sl Phys. Lett.} {\bf #1B} (#2) #3}

\def\AoP#1#2#3{{\sl Ann. of Phys.} {\bf #1} (#2) #3}

\def\PR#1#2#3{{\sl Phys. Reports} {\bf #1} (#2) #3}

\def\IJMPA#1#2#3{{\sl Int. J. Mod. Phys.} {\bf A#1} (#2) #3}

\def\JPA#1#2#3{{\sl J. Physics} {\bf A#1} (#2) #3}

\def\MPLA#1#2#3{{\sl Mod. Phys. Lett.} {\bf A#1} (#2) #3}
\def\JETP#1#2#3{{\sl Sov. Phys. JETP} {\bf #1} (#2) #3}
\def\JETPL#1#2#3{{\sl Sov. Phys. JETP Lett.} {\bf #1} (#2) #3}

\def\JETPL#1#2#3{{\sl  Sov. Phys. JETP Lett.} {\bf #1} (#2) #3}

\def\hepth#1{[{\tt arXiv:hep\--th/#1}\/]}

\def\hepthP#1{{\tt arXiv:hep\--th/#1}}
\def\nlinP#1{{\tt arXiv:nlin.si/#1}}

\def\JHEP#1#2#3{{\sl JHEP} {\bf #1} (#2) #3}
\def\EJPC#1#2#3{{\sl Eur. Phys. J. C} {\bf #1} (#2) #3}
\def\ex#1{{\rm e}^{#1}}
\def\Buildrel#1\over#2\under#3{\mathrel{\mathop{\kern0pt
#2}\limits^{#1}_{#3}}}
\def\alv{\vec{\alpha}}

\def\lav{\vec{\lambda}}
\newcommand\blank[1]{#1}
\renewcommand\blank[1]{}
\begin{document}

\begin{titlepage}
\vspace{-1cm}
\noindent
\vspace{0.0cm}
\hfill{hep-th/0408119}\\
\vspace{0.0cm}
\hfill{August 2004}\\
\phantom{bla}
\vfill
\begin{center}
{\large\bf T-duality in Massive Integrable Field Theories:}
\par \vskip .1in
{\large\bf The Homogeneous and Complex sine-Gordon
Models}
\end{center}

\vspace{0.3cm}
\begin{center}
J. Luis Miramontes
\par \vskip .1in \noindent
{\em 
Departamento de F\'\i sica de Part\'\i culas,\\
and\\
Instituto Gallego de F\'\i sica de Altas Energ\'\i as (IGFAE),\\[5pt]
Facultad de F\'\i sica\\
Universidad de Santiago de Compostela\\
15782 Santiago de Compostela, Spain}\\
\par \vskip .1in \noindent
e-mail: miramont@usc.es
\normalsize
\end{center}
\vspace{.2in}
\begin{abstract}
\vspace{.3 cm}
\small
\par \vskip .1in \noindent

\noindent
The T-duality symmetries of a family of two-dimensional massive integrable
field theories  defined in terms of asymmetric gauged
Wess-Zumino-Novikov-Witten actions modified by a potential are
investigated. These theories are examples of massive non-linear sigma
models and, in general, T-duality relates two different dual sigma models
perturbed by the same potential. When the unperturbed
theory is self-dual, the duality transformation relates two
perturbations of the same sigma model involving different
potentials. Examples of this type are provided by the Homogeneous
sine-Gordon theories, associated with cosets of the form $G/U(1)^r$
where $G$ is a compact simple Lie group of rank $r$. They exhibit a duality
transformation for each element of the Weyl group of $G$ that relates two
different phases of the model. On-shell, T-duality provides a map between the
solutions to the equations of motion of the dual models that changes Noether
soliton charges into topological ones. This map is carefully studied in the
complex sine-Gordon model, where it motivates the construction of
Bogomol'nyi-like bounds for the energy that provide a novel
characterisation of the already known one-solitons solutions where their classical stability becomes explicit.

\end{abstract}

\vfill
\end{titlepage}
\sect{Introduction.}
\label{Intro}

~\indent
It seems difficult to exaggerate the importance of duality in the
investigation of the non-perturbative aspects of quantum field theory,
statistical mechanics, and string theory. In brief, duality symmetries are
discrete transformations that relate either two apparently different
theories, or the same one at different values of its coupling constants.
In some cases, this makes possible to investigate the strong
coupling regime of a given theory from the knowledge of the weak coupling
behaviour of its dual -- and sometimes the original and the dual models
coincide. Concerning two-dimensional non-linear sigma models, the
most characteristic type of duality transformation is target-space
duality, or `T-duality' for short, which is the generalisation of the well
known equivalence of the theories of a single bosonic field compactified on
circles of radii $R$ and $1/R$. Two apparently different
sigma models are said to be T-dual to each other if there is a
canonical transformation between the phase spaces that preserves the
respective Hamiltonians. Due to the important role of
target-space duality in the study of (super) string vacua, there is a lot
known about T-duality for massless sigma models, and we address
the interested reader to~\cite{TDUALITY} where comprehensive reviews and
references to the original literature can be found. In contrast, T-duality
of massive sigma models has received much less attention.

The purpose of this paper is to
discuss T-duality in a particular family of two-dimensional
massive integrable field theories defined  by the gauged
Wess-Zumino-Novikov-Witten (WZW) actions associated with cosets of the form
$G/\bigl(H\times U(1)\bigr)$ modified by a potential. 
It includes
the Homogeneous sine-Gordon (HSG) theories~\cite{HSGClass,HSGQuan1,HSGQuan2},
which correspond to cosets of
the form $G/U(1)^r$ where $G$ is a compact simple Lie group of rank~$r$. They
are generalisations of the complex sine-Gordon
(CSG) model~\cite{LUND,POHL}, which is recovered for
$G=SU(2)$~\cite{BAKAS,PARK}. Other
theories included in this family are the Symmetric-space sine-Gordon
theories constructed in~\cite{HSGClass,SSSG}, and the models
studied by Gomes {\em et al.\/} in~\cite{GOMES1} associated
with cosets of the form $SL(2)\times U(1)^n/U(1)$.

All these two-dimensional theories admit a Lagrangian description in terms of
a massive non-linear sigma model like
\begin{equation} 
{\cal L}= {1\over 2}\> {\cal G}_{ij}(X)\left({\partial
X^i\over\partial t} {\partial X^j\over\partial t} - {\partial
X^i\over\partial x} {\partial X^j\over\partial x}\right) + {\cal
B}_{ij}(X)\> {\partial X^i\over\partial t} {\partial X^j\over\partial x} -
{\cal U}(X)\>,
\lab{SigmaModPot}
\end{equation} 
where $i=1\ldots n$ with $n$ the dimension of the target space, ${\cal G}$ is
a metric, ${\cal B}$ is an antisymmetric tensor, and
${\cal U}$ is a potential. They exhibit a global
$U(1)$ symmetry, which gives rise to an abelian T-duality transformation that
relates off-shell two different massive sigma models associated with the same
coset. An important feature is that the potential
always remains invariant under the
$U(1)$ symmetry and, therefore, the resulting duality transformations are
actually a reflection of the T-duality symmetries of the unperturbed gauged
WZW models, which are well established in the literature~\cite{KIR,KIRplus}. 
In particular, if the coset is of the form $G/U(1)$, T-duality relates two
different Lagrangians involving gauge transformations of vector or axial
form, which is just a consequence of the well known duality between the
$U(1)$ vector-gauged WZW model and the axially-gauged one.

Our discussion will be mostly classical, but our
results are expected to be helpful to elucidate the consequences of duality
in the corresponding quantum theories. At the
quantum level, the formulation in terms of gauged WZW
actions leads to a natural non-perturbative definition of the theory as a
perturbed coset conformal field theory (CFT) specified by an action
of the form~\cite{ZAMO}
\begin{equation} 
S= S_{\rm CFT} + \mu_a \int d^2x\> \Phi^a\>.
\lab{PCFT}
\end{equation} 
Here $S_{\rm CFT}$ denotes an action for the unperturbed
two-dimensional CFT corresponding to the gauged WZW
model, $\Phi^a$
are spinless operators found in the operator algebra of $S_{\rm
CFT}$, and $\mu_a$ are dimensionful real parameters (coupling constants).
The Lagrangian~\rf{SigmaModPot} describes the
theory in the semiclassical limit, so that the massless sigma model ${\cal
L}\big|_{U=0}$ provides a Lagrangian description for the CFT corresponding to
$S_{\rm CFT}$, and the potential becomes identified with the perturbation. In
general, duality transformations relate two different dual CFTs perturbed
by the same potential. However, in some cases the non-linear sigma models
corresponding to the dual CFTs coincide, up to a change of the field
variables. Then, the CFT is said to be self-dual, and the duality
transformation becomes equivalent to a change of the potential or,
equivalently, of the coupling constants $\{\mu_a\}$ in~\rf{PCFT}. This can be
understood as a relationship between different phases of the same perturbed
CFT akin to the Kramers-Wannier duality between the high and low temperature
phases of the Ising model~\cite{KW}. Recall that the scaling
$T\ra T_c$ limit of the Ising model is described by an action of the
form~\rf{PCFT} where the unperturbed CFT is the critical Ising model, and the
perturbation is defined by a single operator (the thermal or energy-density
operator), such that
$\mu\sim T-T_c$. Then, the duality between the high ($T>T_c$) and low
($T<T_c$) temperature phases corresponds to the duality of the
perturbed CFT theory under $\mu \ra -\mu$.

The paper is organised as follows. In section~\ref{BuscherPot} we
provide a  brief description of the duality transformations exhibited by the
family of integrable theories using
Buscher's formulation of abelian T-duality. In section~\ref{Models}, we
introduce the particular class of integrable theories. In
section~\ref{OffShell}, we explicitly construct the canonical
transformation that relates a given theory to its dual. In
section~\ref{OnShell}, the duality transformation is considered on-shell,
which provides a map between the solutions to the equations of motion of the
two dual massive sigma models that interchanges Noether and topological
charges. Finally, in sections~\ref{Examples} and~\ref{CSG}, we illustrate our
results with a number of examples associated with cosets of the form
$G/U(1)^p$. In particular, we discuss T-duality in the HSG and the CSG
models. In the HSG models, we also characterise their phase structure to show
that T-duality actually provides a relation between different phases.
In the CSG model, our discussion motivates the construction of
Bogomol'nyi-like bounds for the energy saturated by the already
known solitons which, to our knowledge, have not been reported before in
the literature.

\sect{Potentials and T-duality.}
\label{BuscherPot} 

~\indent
A useful description of abelian T-duality in the massless case was
introduced originally by Buscher~\cite{BUSCHER,TDUALITY}. It can be
summarised as follows. Consider the $1+1$ dimensional bosonic
non-linear sigma model defined by the Lagrangian~\rf{SigmaModPot} with
${\cal U}=0$. Assume that the sigma model has an abelian isometry and
that we have chosen coordinates adapted to the isometry such that it is
represented simply by a translation in the coordinate $X^1$, which
requires that ${\cal G}$ and ${\cal B}$ are independent of
$X^1$. Then, T-duality is a transformation that relates the non-linear sigma
model corresponding to $({\cal G},{\cal B})$ to another one specified by
\begin{eqnarray}
&&{\cal G}^{\rm D}_{11}={1\over {\cal G}_{11}}\>, \qquad {\cal G}^{\rm D}_{1i}
={{\cal B}_{1i}\over {\cal G}_{11}}\>, \qquad {\cal G}^{\rm D}_{ij}= {\cal
G}_{ij} -{{\cal G}_{1i}{\cal G}_{1j}-{\cal B}_{1i}{\cal B}_{1j} \over {\cal
G}_{11}}\>,\nn[5pt]
&&{\cal B}^{\rm D}_{1i}={{\cal G}_{1i}\over {\cal
G}_{11}}\>,
\qquad {\cal B}^{\rm D}_{ij}= {\cal B}_{ij} -{{\cal G}_{1i}{\cal
B}_{1j}-{\cal B}_{1i}{\cal G}_{1j} \over {\cal G}_{11}}\>, \qquad\quad
i,j\not=1\>;
\lab{Buscher}
\end{eqnarray}
moreover, $({\cal G}^{\rm D})^{\rm D}={\cal G}$ and
$({\cal B}^{\rm D})^{\rm D}={\cal B}$. Both sigma models are related by a
canonical transformation between the phase spaces that preserves the
respective Hamiltonians~\cite{AGcan,VENEZCan}. Consequently, even though
they are generally defined by completely different Lagrangians, the
sigma models specified by $({\cal G},{\cal B})$ and $({\cal G}^{\rm
D},{\cal B}^{\rm D})$ describe the same physics. 

In the massive case,
conformal invariance is explicitly violated by the presence of the
potential ${\cal U}$. Then, a general description of T-duality is not
available, but we shall study a family of theories where
the duality properties directly follow from those of massless sigma
models. Suppose that 
${\cal U}(X)$ is a function of the coordinates and not
of their derivatives, and that it preserves the abelian isometry,
so that ${\cal U}(X)$ is also independent of $X^1$. Then, the potential
does not change under the canonical transformation corresponding
to~\rf{Buscher} and T-duality relates the massive models specified by
$({\cal G},{\cal B},{\cal U})$ and
$({\cal G}^{\rm D},{\cal B}^{\rm D},{\cal U})$. This will be indicated as
follows
\begin{equation}
({\cal G},{\cal B},{\cal U}) 
\buildrel {\rm T-duality}\over{\hbox to
65pt{\rightarrowfill}} ({\cal G}^{\rm D},{\cal B}^{\rm D},{\cal U}),
\lab{TDualMass}
\end{equation}
which provides a relationship between two {\em different} dual sigma 
models perturbed by the same potential.
Explicit examples of massive integrable sigma models related by duality
transformations like~\rf{TDualMass} can be found
in~\cite{GOMES1,GOMES2}.

A interesting particular case occurs when the unperturbed non-linear
sigma model is `self-dual'.
By this we mean that there is a change of field variables
$X^i \rightarrow
\widetilde{X}^i$ (a point transformation in the context of canonical
transformations) that provides a relationship of the form
$({\cal G}^{\rm D},{\cal B}^{\rm D},{\cal U}) \rightarrow  ({\cal G}, {\cal
B},\widetilde{{\cal U}})$. Then, eq.~\rf{TDualMass} becomes
\begin{equation}
({\cal G},{\cal B},{\cal U}) 
\buildrel {\rm T-duality} \over{\hbox to
60pt{\rightarrowfill}}
({\cal G}^{\rm D},{\cal B}^{\rm D},{\cal U})
\buildrel {X^i \rightarrow \widetilde{X}^i} \over{\hbox to
55pt{\rightarrowfill}}
({\cal G},{\cal B},\widetilde{{\cal U}})\>,
\lab{TPoint}
\end{equation}
which relates two perturbations of the {\em same} non-linear sigma model
by different potentials. The relevance of~\rf{TPoint} becomes clearer
when considering the non-perturbative definition of the model at the
quantum level by means of an action of the form~\rf{PCFT}. Then, the
potential is identified with the perturbation, and eq.~\rf{TPoint} can be
understood as a duality symmetry between two different phases of the same
model characterised by the domain where the coupling constants $\mu_a$ take
values. The HSG and CSG models, which will be discussed in
sections~\ref{Examples} and~\ref{CSG}, provide examples where T-duality
is of the form~\rf{TPoint}. 

\sect{The integrable theories.}
\label{Models} 

~\indent
Eqs.~\rf{TDualMass} and~\rf{TPoint} summarise the form of the
abelian T-duality transformations exhibited by a large family of
two-dimensional integrable theories. 
They are particular cases of an even
larger class of integrable models associated with generic cosets
constructed in the literature by several authors with different
purposes. Namely, description of deformed coset
models~\cite{PARK,PARKSHIN}, Hamiltonian reduction of two-loop WZW
models~\cite{GOMES1,GOMES2,TLWZW}, Lagrangian formulation of reduced
symmetric space sigma models~\cite{SSSM}, or, like
in~\cite{HSGClass,MASSIVE}, simply as the Lagrangian models whose
classical equations of motion are the non-abelian affine Toda (NAAT)
equations of Leznov and Saveliev~\cite{NAAT}, which can be recognised
as their distinctive common feature.
Following~\cite{SSSM}, these theories can be defined in a systematic way
in terms of a triplet of Lie algebras $({\go h},{\go g},{\go f})$, with
respective associated Lie groups $H\subset G\subseteq F$, as follows -- in
the following, we will always assume that $F$ is simple, although the
construction is not restricted to this case. If
$G\not=F$, we shall assume that $F/G$ is a symmetric space, which means
that there is a Lie algebra decomposition ${\go f}={\go g}\oplus {\go k}$
that satisfies
\begin{equation}
[{\go g},{\go g}]\subseteq {\go g} \>, \qquad [{\go g},{\go k}]\subseteq 
{\go k} \>, 
\qquad [{\go k},{\go k}]\subset {\go g}\>.
\end{equation}
Then, we choose two arbitrary (adjoint-diagonalisable) constant elements
$T_+$ and $T_-$ in~${\go k}$. Correspondingly, if $G=F$ we choose $T_\pm$ in ${\go g} = {\go
f}$.\foot{This case can equivalently be described in terms of the
symmetric space $G\times G/G$ of type~II in Cartan's
classification.} Finally,
${\go h}$ is defined as the simultaneous centraliser of $T_+$ and $T_-$ in
${\go g}$, namely 
${\go h}=\{u\in {\go g} \mid [u,T_+]= [u,T_-]= 0\}$, and the model is
defined by the action~\cite{HSGClass}
\begin{eqnarray}
&&S^{\{\tau\}}[h,A_\pm]= k\left( S_{\rm gWZW}^{\{\tau\}}[h,A_\pm] -  \int
d^2 x\>  U(h)\right)
\lab{ModAction}\\ 
&& S_{\rm gWZW}^{\{\tau\}}[h,A_\pm]= S_{\rm
WZW}[h] +{1\over\pi} \int d^2x\> \Bigl(-\bil{A_+}{\partial_-h
h^{-1}} \nn 
&& \hskip 2.5truecm + \bil{\tau(A_-)}{h^{-1}\partial_+ h} +
\bil{h^{-1}A_+h}{\tau(A_-)} - \bil{A_+}{A_-} \Bigr)\>,
\lab{gWZW}\\
&& U(h)= \lambda\> \bil{T_+}{h^{-1}\>T_- h}\>.
\lab{Potential}
\end{eqnarray}
Here, $kS_{\rm WZW}[h]$ is the usual WZW action at level $k$ for the
bosonic field
$h$ taking values in (some faithful representation of) $G\subseteq F$,
$A_\pm$ are non-dynamical gauge fields taking values in ${\go h}$,
$\lambda$ is a dimensionful parameter,
$\bil{\>}{\>}$ is the invariant bilinear form of ${\go f}$
normalised such that $S_{\rm WZW}[h]$ is defined modulo $2\pi{\fl
Z}$~\cite{WZW}, and we are using the notation
$\partial_\pm = \partial/\partial x_\pm$ with $x_\pm=t\pm x$. In this
action, $(\lambda,T_+,T_-)$ play the role of coupling constants.

The action $S_{\rm gWZW}^{\{\tau\}}$ is invariant under the
group of gauge transformations
\begin{equation}
h\rightarrow \alpha\> h\> \widehat{\tau}(\alpha^{-1})\>, \qquad
A_\pm \rightarrow \alpha \> A_\pm\> \alpha^{-1}- \partial_\pm\alpha\>
\alpha^{-1}\>,
\lab{GaugeT}
\end{equation}
where $\alpha=\alpha(t,x)$ takes values in $H$, and $\widehat\tau$ is the
lift of a suitable automorphism $\tau$ of
${\go h}$ that leaves the restriction of the bilinear form
$\bil{\>}{\>}$  to ${\go h}$ invariant so that the
gauge group is `anomaly free'~\cite{HSGClass,GWZW}. The lift $\widehat\tau$ is
defined by $\widehat\tau\bigl(\ex{u}\bigr) = \ex{\tau(u)}$, for any $u\in{\go
h}\>$. In particular,  taking
$\tau=+I$ or~$-I$ leads to gauge transformations of vector or axial
type, respectively. Then, $k S_{\rm gWZW}^{\{\tau\}}$ is the action of
a 
$G/H$ coset conformal field theory (CFT) at level $k$~\cite{GWZW}. The
particular coset model depends on $\tau$. The usual one is
recovered with $\tau=+I$, which is the only case considered
in~\cite{SSSM}, while for
$\tau\not=I$ the models are examples of the asymmetric coset models
constructed in~\cite{ASYM}. Taking the relationship
between the constant elements $T_\pm$ and $H$ into account, it is easy
to check that
$U(\alpha h \beta)= U(h)$ for any $\alpha$, $\beta$ in $H$,
which shows that the potential is uniquely defined on the coset $G/H$
independently of the choice of $\tau$.

The connection of these models with the NAAT equations is obtained as follows.
The variation of the action~\rf{ModAction} with respect to the field~$h$
yields the equations of motion that can be expressed in a zero-curvature
form~\cite{HSGClass}
\begin{equation}
\Bigl[ \partial_+  + h^{-1}\>\partial_+h  + h^{-1} 
A_+ h - \pi\lambda\> \xi\> T_+ \>, \> \partial_-  +
\tau(A_-)+ \xi^{-1}\> h^{-1}
\>T_- h  \Bigr] =0\>,
\lab{GNaT}
\end{equation}
where $\xi$ is a constant spectral parameter. Correspondingly, the
variations with respect to
$A_\pm$ lead to the  constraints
\begin{eqnarray}
&& P_{\go h}\Bigl( h^{-1}\>\partial_+h  + h^{-1} A_+
h\Bigr) - \tau(A_+) = 0\>,\nn
&& P_{\go h}\Bigl( -\partial_- h\> h^{-1} + h\> \tau(A_-)
h^{-1} \Bigr) - A_- = 0\>,
\lab{GConst}
\end{eqnarray}
where $P_{\go h}$ is the projector onto the subalgebra ${\go h}$.
Next, projecting~\rf{GNaT} onto ${\go h}$ and using~\rf{GConst}, it can be
checked that the gauge field is flat on-shell: $[\partial_+ \>+ \>A_+\>,
\>\partial_- 
\>+ \>A_-]=0$. Since we are considering theories defined on
two-dimensional Minkowski space, this allows one to fix the gauge by
setting
$A_+=A_-=0$. In this gauge, the equations of motion~\rf{GNaT} and the
constraints~\rf{GConst} simplify to
\begin{eqnarray}
&& \partial_-\bigl(h^{-1}\partial_+h\bigr) = -\pi\lambda\>
\Bigl[T_+,h^{-1}T_- h\Bigr]\>, \lab{NaaT}\nn
&& P_{\go h}\Bigl( h^{-1}\>\partial_+h \Bigr)= P_{\go h}\Bigl( \partial_-
h\> h^{-1}\Bigr)= 0\>, \lab{ConstNaaT}
\end{eqnarray}
which is a system of
non-abelian affine Toda equations~\cite{NAAT,HSGClass,TLWZW}.

Notice that~\rf{gWZW} is quadratic in the non-dynamical
gauge fields. Therefore, they can be
integrated out by solving~\rf{GConst} for $A_\pm$ to obtain a sigma
model description of the coset CFT and, hence, of the massive theory in
terms of a Lagrangian like~\rf{SigmaModPot}. 
However, in order to leave
the choice of the gauge fixing prescription completely free, we will keep
the gauge fields and work directly with~\rf{ModAction}

Since we are interested in abelian T-duality, we will restrict ourselves to
cases where $H$ contains, at least, a $U(1)$ factor; {\em i.e.\/},
$H$ will be of the form $\widehat{H}\times U(1)$. If we call $T^0$ the
generator of the $U(1)$ factor, we can choose a basis of generators
$\{t^a,T^0\}$ for the Lie algebra ${\go h}=\widehat{\go h}\oplus u(1)$ such
that
\begin{equation}
\bil{T^0}{T^0} =-1\quad \text{and}\quad  \bil{t^a}{T^0} =0\quad \forall~a\>.
\end{equation}
In this basis, the gauge fields split in components as follows
\begin{equation}
A_\pm = A_\pm^a t^a + a_\pm T^0\equiv \widehat{A}_\pm  + a_\pm T^0\>.
\end{equation}
We will also assume that the $U(1)$ factor is compact and
${T^0}^\dagger=-T^0$. Then, the WZW field can be parametrised as 
\begin{equation}
h=\ex{\>\beta\> T^0}\> h_0\>
\ex{\>\gamma\> \tau(T^0)}\>,
\lab{WZWfield}
\end{equation}
with $\beta$ and $\gamma$ real bosonic fields, so that the
$\tau$-dependent
$U(1)$ gauge transformations~\rf{GaugeT} generated by the elements of
$H$ of the form $\alpha=\ex{\>\rho\> T_0}$ correspond to
\begin{equation}
\beta\rightarrow \beta+ \rho\>, \qquad \gamma\rightarrow \gamma-\rho\>,
\qquad a_\pm \rightarrow a_\pm -\partial_\pm \rho\>,
\lab{GaugeSimple}
\end{equation}
while $h_0$ remains invariant. This way, and using the Polyakov-Wiegmann
formula
\begin{equation}
kS_{\rm WZW}[gh] = kS_{\rm WZW}[g] + kS_{\rm WZW}[h] -{k\over\pi} \int
d^2x\> \bil{g^{-1}\partial_+ g}{\partial_- h h^{-1}}\>,
\lab{PW}
\end{equation}
the action~\rf{gWZW} can be written in the form\foot{It is well known
that the Polyakov-Wiegmann formula only holds if the WZW model is
defined on a simply connected compact manifold; otherwise, it has to
be corrected by adding topological terms~\cite{PWTOP}. In our case, we
have used this freedom to remove a term of the form 
$
-{k\over2\pi}\int
d^2 x\> \Bigl(\partial_+(\beta \partial_-\gamma) - \partial_-(\beta
\partial_+\gamma)\Bigr)
$ 
in order to ensure explicit gauge invariance.}
\begin{eqnarray}
&& S^{\{\tau\}}[\ex{\>\beta\> T^0}\> h_0\>
\ex{\>\gamma\> \tau(T^0)}, A_\pm] =\nn
&&  \hskip1truecm ={k\over2\pi}\int
d^2 x\> \left(- \partial_+\phi \partial_-\phi +
2\widetilde{a}_+ \partial_-\phi -2\widetilde{a}_- \partial_+\phi 
\right) \>+\> \Delta S^{\{\tau\}}[ h_0, A_\pm]\>,
\lab{ModActionPlus}
\end{eqnarray}
where
\begin{eqnarray}
&&\Delta S^{\{\tau\}}[ h_0, A_\pm] = k\left( S_{\rm
gWZW}^{\{\tau\}}[h_0,\widehat{A}_\pm] - 
\int d^2 x\>  U(h_0)\right)\nn
&&  \hskip1truecm + {k\over\pi}\Bigl(\bigl(1+R^\tau(h_0)\bigr)\>
\widetilde{a}_+
\widetilde{a}_- -\widetilde{a}_+ J_-(h_0) + \widetilde{a}_-
J_+^{\tau}(h_0) \nn &&  \hskip2truecm +\bil{\widehat{A}_+}{
h_0\> \tau(T_0)\> h_0^{-1}}\>
\widetilde{a}_- + \bil{\tau(\widehat{A}_-)}{h_0\> T_0\> h_0^{-1}}
\>\widetilde{a}_+ \Bigr)\>,
\lab{ModActionPlusB}
\end{eqnarray}
and we have introduced the gauge invariant fields
(see~\rf{GaugeSimple})
\begin{equation}
\phi =\beta+ \gamma\>, \qquad \widetilde{a}_+ =a_+ + \partial_+ \beta\>,
\qquad \widetilde{a}_- =a_- - \partial_- \gamma\>, 
\end{equation}
together with
\begin{eqnarray}
&R^\tau(h_0) = \bil{\tau(T_0)}{h_0^{-1}\> T_0\> h_0}& \nn[5pt]
&J_-(h_0) = \bil{T_0}{\partial_- h_0 h_0^{-1}} \>, \quad 
J_+^\tau(h_0) = \bil{\tau(T_0)}{h_0^{-1} \partial_+
h_0}\>.&
\lab{Definitions}
\end{eqnarray}

\sect{Off-shell T-duality.}
\label{OffShell}

~\indent
The action~(\ref{eq:ModAction}) or,
equivalently,~(\ref{eq:ModActionPlus}) is invariant under the global
$U(1)$ transformation
\begin{equation}
h \rightarrow \ex{\>\rho\> T^0}\> h\>
\ex{\>\rho\> \tau(T^0)}\>,
\lab{Global}
\end{equation}
which corresponds to $\phi(t,x)\rightarrow \phi(t,x)+2\rho$ and, therefore,
$\phi$ is an adapted coordinate for this symmetry transformation. 
As explained in~\cite{AGcan}, in order to find the dual action
associated with the abelian isometry~\rf{Global} by means of a canonical
transformation we can use the Routh function with respect to
$\phi$, which means that the Legendre transformation is only performed with
respect to this coordinate. Let us write
$S^{\{\tau\}}[h,A_\pm]= \int d^2x\> {\cal
L}^{\{\tau\}}[h,A_\pm]$. Taking~\rf{ModActionPlus} into account, the
relevant canonical momentum is
\begin{equation}
\Pi_\phi= {\partial {\cal L}^{\{\tau\}} \over \partial (\partial_t \phi)}
= {k\over 4\pi}\Bigl(-\partial_t \phi +2(\widetilde{a}_+ -
\widetilde{a}_-)\Bigr)\>,
\end{equation}
and the Routh function is given by
\begin{eqnarray}
&&{\cal R}^{\{\tau\}}(\phi,\Pi_\phi) = \partial_t\phi\>
\Pi_\phi- {\cal L}^{\{\tau\}}\nn
&& \hskip0.5truecm =\> -{2\pi\over k}\Pi_\phi^2 + 2(\widetilde{a}_+-
\widetilde{a}_-)\Pi_\phi 
-{k\over8\pi}(\partial_x \phi)^2
+ {k\over2\pi}(\widetilde{a}_+ + \widetilde{a}_-) \partial_x \phi\nn
&& \hskip1.5truecm   -{k\over 2\pi} (\widetilde{a}_+-
\widetilde{a}_-)^2 - \Delta{\cal L}^{\{\tau\}}\>,
\end{eqnarray}
where $\Delta{\cal L}^{\{\tau\}}$ is the piece of ${\cal L}^{\{\tau\}}$
corresponding to~\rf{ModActionPlusB}. Now, following~\cite{AGcan}, we consider
the canonical transformation generated by
\begin{equation}
F=-{k\over 8\pi}\int_{-\infty}^{+\infty} dx\> \left(\partial_x\phi\>
\phi^{\rm D} -\phi\> \partial_x\phi^{\rm D}\right)\>;
\end{equation}
namely,\foot{This canonical transformation corresponds to a duality
transformation of the form~\rf{Buscher} together with a trivial change of the
normalisation of the dual field $\phi^{\rm D}\rightarrow
-(k/4\pi)\> \phi^{\rm D} $.}
\begin{equation}
\Pi_\phi = {\delta F\over \delta \phi}={k\over4\pi}\>
\partial_x\phi^{\rm D}\>,
\qquad
\Pi_{\phi^{\rm D}} = -{\delta F\over \delta \phi^{\rm D}}= 
{k\over4\pi}\>\partial_x\phi\>.
\lab{CanTransf}
\end{equation}
This transformation preserves the Hamiltonian and, hence, the
dual Routh function is obtained just by making these substitutions in
${\cal R}^{\{\tau\}}$,
\begin{eqnarray}
&&{\cal R}^{\{\tau\}}(\phi,\Pi_\phi) =
{\cal R}^{\{\tau\}\>{\rm D}}(\phi^{\rm D},\Pi_{\phi^{\rm
D}}) \nn[5pt]
&& \hskip0.5truecm =\> -{2\pi\over k}\Pi_{\phi^{\rm
D}}^2 + 2(\widetilde{a}_+ +
\widetilde{a}_-)\Pi_{\phi^{\rm D}} 
-{k\over8\pi}(\partial_x \phi^{\rm D})^2
+ {k\over2\pi}(\widetilde{a}_+ - \widetilde{a}_-) \partial_x \phi^{\rm
D}\nn 
&& \hskip1.5truecm   -{k\over 2\pi} (\widetilde{a}_+-
\widetilde{a}_-)^2 - \Delta{\cal L}^{\{\tau\}}\>.
\lab{DualRouth}
\end{eqnarray}
Performing the inverse Legendre
transform, it is not difficult to check that~\rf{DualRouth} corresponds to
the dual action
\begin{equation}
S^{\{\tau\}\>{\rm D}}[\ex{\>\beta\> T^0}\> h_0\>
\ex{\>\gamma\> \tau(T^0)}, A_+,A_-] =
S^{\{\tau\cdot \sigma_{T_0}\}}[\ex{\>\beta\> T^0}\> h_0\>
\ex{\>-\gamma\> \tau (T^0)}, A_+,A_-^{'}]\>,
\lab{U1Dual}
\end{equation}
where $A_-^{'}= \sigma_{T_0}(A_-) + 2\partial_- \gamma\> T^0$, and
$\sigma_{T_0}$ is the following involutive automorphism of ${\go h}$:
\begin{equation}
\sigma_{T_0}(u) = u - 2\; {\bil{u}{T_0}\over \bil{T_0}{T_0}}\; T_0\>, \qquad
\forall u\in{\go h}\>;
\lab{Reflect}
\end{equation}
{\em i.e.\/}, $\sigma_{T_0}$ is a reflection on the subspace of ${\go h}$
orthogonal to
$T_0$, and $\sigma_{T_0}^2=I$.
Therefore, up to the trivial field transformations
\begin{equation}
h=\ex{\>\beta\> T^0}\> h_0\> \ex{\>\gamma\> \tau(T^0)}
\longrightarrow
h'=\ex{\>\beta\> T^0}\> h_0\> \ex{\>-\gamma\> \tau (T^0)}\>,
\qquad A_- \rightarrow A_-^{'}\>,
\end{equation}
the only effect of the T-duality
transformation associated with the $U(1)$ transformation~\rf{Global}
is to change the action corresponding to the automorphism $\tau$
into the action corresponding to $\tau\cdot\sigma_{T_0}$,
\begin{equation}
S^{\{\tau\}}
\buildrel {\rm T-duality}\over{\hbox to
65pt{\rightarrowfill}} S^{\{\tau\cdot \sigma_{T_0}\}}
\lab{U1DualAct}
\end{equation}
Notice that the potential~\rf{Potential} remains
invariant and, hence, the duality
transformation summarised by~(\ref{eq:U1Dual}) and~(\ref{eq:U1DualAct})
is precisely of the form~\rf{TDualMass}.

\sect{On-shell T-duality.}
\label{OnShell} 

~\indent
On-shell, eq.~\rf{CanTransf} provides a map
between the solutions to the equations of motion of the two dual massive
sigma models, which can be characterised as a {\em pseudoduality}
transformation using the terminology of~\cite{ORLANDOPseudo}. In  the
string theory context, it is well known that T-duality changes trivial
boundary conditions (momentum modes) into non-trivial ones (winding
modes). In our case, even though we consider theories defined on
Minkowski space where winding numbers are not defined, the on-shell
T-duality transformations will admit a similar
interpretation when restricted to soliton solutions, changing Noether
(electric) soliton charges into topological (magnetic) ones. 

The (gauge invariant)
solutions to the equations of motion of~\rf{ModActionPlus}
will be specified by $(h_0,\phi)$. This way, the map provided
by~\rf{CanTransf} corresponds to $(h_0,\phi)\rightarrow (h_0,\phi^{\rm
D})$, where $\phi^{\rm D}$ is the solution to the equations of the
canonical transformation understood as partial differential equations for
$\phi^{\rm D}$:
\begin{equation}
\partial_x\phi^{\rm D} = {(1-R^\tau) \partial_t
\phi -2 ({\cal J}_+^\tau +
{\cal J}_-^\tau) \over 1+R^\tau}\>, \qquad
\partial_t\phi^{\rm D} =
{(1-R^\tau) \partial_x
\phi -2 ({\cal J}_+^\tau -
{\cal J}_-^\tau) \over 1+R^\tau}\>,
\lab{Pseudo}
\end{equation}
where
\begin{equation}
{\cal J}_+^\tau = J_+^\tau(h_0) + \bil{\widehat{A}_+}{h_0\> \tau(T^0)\>
h_0^{-1}}\>, \qquad
{\cal J}_-^\tau = J_-(h_0) -\bil{\tau(\widehat{A}_-)}{h_0\> T^0\>
h_0^{-1}}\>,
\end{equation}
and $R^\tau=R^\tau(h_0)$, $J_+^\tau(h_0)$ and $J_-(h_0)$ have been
previously defined in~\rf{Definitions}. The $U(1)$ global
symmetry~\rf{Global} leads to the conserved Noether current $J^{{\rm
N}\>\mu} = 2\>\partial {\cal L}^{\{\tau\}}/\partial(\partial_\mu\phi)$,
and the eqs.~\rf{Pseudo} can be written as
\begin{equation}
\partial_x\phi^{\rm D} = {2\pi\over k}\> J^{\rm N}_0\>, \qquad
\partial_t\phi^{\rm D} ={2\pi\over k}\> J^{\rm N}_1\>,
\lab{PseudoNoeth}
\end{equation}
whose integrability is equivalent to the
conservation of $J^{{\rm N}\>\mu}$,
which holds on-shell.

Consider a soliton solution to the equations of motion of
$S^{\{\tau\}}$ that carries a definite value of the $U(1)$ Noether charge
$Q^{\rm N}=\int_{-\infty}^{+\infty} dx\> J_0^{\rm N}$.
Integrating the equations~\rf{PseudoNoeth}, we obtain
\begin{eqnarray}
&&\phi^{\rm D}(t,+\infty)- \phi^{\rm D}(t,-\infty)=
{2\pi\over k} \int_{-\infty}^{+\infty}
dx\> J_0^{\rm N} = {2\pi\over k}Q^{\rm N}\nn
&&Q^{\rm (D)\> N}=\int_{-\infty}^{+\infty}
dx\> J_0^{\rm (D)\> N} = {k\over 2\pi}\left(\phi(t,+\infty)-
\phi(t,-\infty)\right)\>.
\lab{ExplicitInt}
\end{eqnarray}
It is important to notice that, due to the continuous $U(1)$ symmetry of
the potential, the values of the boundary conditions $\phi(t,\pm\infty)$
are not quantised. This means that, in general, their value will change
continuously with~$t$ and, thus, they do not provide proper conserved
topological charges which can be used to classify the solutions of
the equations of motion. In contrast, for each pair of dual soliton
solutions, and since the Noether charges
$Q^{\rm N}$ and
$Q^{\rm (D)\> N}$ are conserved, the eqs.~\rf{ExplicitInt}
show that
$\omega= \phi(t,+\infty)-
\phi(t,-\infty)$ is indeed time independent. In other words, $w$ provides a
conserved quantity, but its conservation is not of topological
nature; it follows from the equations of motion in the dual phase. With this
caveat,
$\omega$ can be understood as a topological charge carried by the soliton
solutions associated with the current
\begin{equation}
J_\mu^{\rm T} = - \varepsilon_{\mu\nu}\>\partial^\nu \phi\>,
\qquad
\varepsilon_{01}=+1,
\lab{TopCurr}
\end{equation}
so that $Q^{\rm T}=\int_{-\infty}^{+\infty} dx\> J_0^{\rm
T}=\phi(t,+\infty)- \phi(t,-\infty)=\omega$. This way, soliton solutions
can be specified by
$[\omega,Q^{\rm N}]$, where $\omega$ and $Q^{\rm N}$ are constant,
and~\rf{ExplicitInt} shows that the solutions to the  equations of
motion of
$S^{\{\tau\}}$ characterised by
$[\omega,(k/2\pi)\omega^{\rm D}]$ are T-dual to those of
$S^{\{\tau\cdot\sigma_{T_0}\}}$  labelled by
$[\omega^{\rm D},(k/2\pi)\omega]$.

Recall now
that $\phi$ is a compact field, of angular nature, which is identified with
$\phi +2\pi\Delta$ for some constant real number
$\Delta$ that depends on the normalisation of $\phi$. Consequently, the
topological charges $\omega$ and $\omega^{\rm D}$ are defined modulo
$2\pi\Delta$ and 
$2\pi\Delta^{\rm D}$, respectively, and 
T-duality implies that the value of the Noether charge $Q^{\rm N}$ is
uniquely defined only modulo $k\Delta^{\rm D}$.

In contrast to the semiclassical description which has been considered so
far, under quantisation the Noether charges carried by the solitons become
quantised in terms of some unit charges $q$ and
$q^{\rm D}$; {\em i.e.\/},
$Q^{\rm N}\in q\>{\fl Z}$ and
$Q^{\rm (D)\> N}\in q^{\rm D}\>{\fl Z}$. This, together with the
identification
$Q^{\rm N}\sim Q^{\rm N}+ k\Delta^{\rm D}$, can be seen as an indication
that the global
$U(1)$ symmetry will be broken to a discrete symmetry associated with a
finite subgroup of $U(1)$ characterised by $k$ -- {\em e.g.\/},
${\fl Z}_k$ --, which is actually expected to occur in the quantum theory.
Moreover, through the duality transformation, the boundary
conditions become quantised too: $\omega\in (2\pi q^{\rm D}/k)\> {\fl Z}$
and $\omega^{\rm D}\in
(2\pi q/k)\> {\fl Z}$, which together with 
$\omega \sim \omega+ 2\pi\Delta$ indicates that the
solitonic quantum field configurations will be parafermionic.

\sect{Examples.}
\label{Examples} 

~\indent
We now describe T-duality in some specific
models associated with cosets of the form $G/U(1)^p$, with $p\geq1$.

\subsection{{\mathversion{bold}$G/U(1)$} models.}
\label{VectorAxial} 

~\indent
The simplest examples where T-duality is described by~\rf{U1DualAct}
are provided by the models corresponding to cosets of the form
$G/U(1)$. In particular, they include the Complex
sine-Gordon theory, which will be discussed in detail in
section~\ref{CSG}, and the models studied by Gomes {\em et al.\/}
in~\cite{GOMES1}. They are associated with
$SU(2)/U(1)$ and with cosets of the form
$SL(2)\times U(1)^n/U(1)$, respectively. In this case, there are only two
possibilities for the automorphism $\tau$. It can be either
$\tau=+I$ or $\tau=-I$, which lead to $U(1)$ gauge transformations of
vector or axial form. Correspondingly, the automorphism
defined in~\rf{Reflect} is $\sigma_{T_0}=-I$ and, hence, T-duality is
simply a reflection of the well known fact that the
$U(1)$ vector-gauged WZW model is dual to the axially-gauged
one~\cite{KIR}. 

\subsection{{\mathversion{bold}$G/U(1)^p$} models with {\mathversion{bold}$p>1$}: T-duality in
the HSG models.}
\label{HSG}

~\indent
Naively, when $H$ contains more than one $U(1)$ factor,  each $T^0\in {\go h}$
gives rise to a T-duality transformation of the form~\rf{U1DualAct}, which
will be denoted $D_{T^0}$. These transformations can be multiplied. The
product of two of them can be defined simply as the result of
performing one after the other, so that
$D_{T^0} D_{V^0}$ is specified by the diagram
\newcommand{\sts}{\footnotesize}
\setlength{\unitlength}{1mm}
\newsavebox{\Dual}
\sbox{\Dual}{\begin{picture}(85,5)(10,-15)
\put(10,-1){$S^{\{\tau\}}$}
\put(19,0){\vector(1,0){20}}
\put(25,3){\sts$D_{V^0}$}
\put(41,-1){$S^{\{\tau\cdot \sigma_{V^0}\}}$}
\put(56,0){\vector(1,0){20}}
\put(62,3){\sts$D_{T^0}$}
\put(78,-1){$S^{\{\tau\cdot \sigma_{V^0}\cdot
\sigma_{T^0}\}}$}
\put(11,-15){\line(0,1){11}}
\put(11,-15){\line(1,0){69}}
\put(40,-12){\sts $D_{T^0}
D_{V^0}$}
\put(80,-15){\vector(0,1){11}}
\end{picture}}
\begin{equation}
\begin{picture}(130,25)(-15,-19)
\put(10,-15){\usebox{\Dual}}
\end{picture}
\lab{Product}
\end{equation}
which relates the models corresponding to $\tau$ and $\tau\cdot
\sigma_{V^0}\cdot \sigma_{T^0}$ without changing the potential. The resulting
set of duality transformations forms a non-abelian group where
$D_{T^0} D_{T^0} =I$ and
$D_{T^0} D_{V^0} \not= D_{V^0} D_{T^0} $ unless $\bil{T^0}{V^0}=0$.  
However,  already at the classical level, not all the resulting
transformations are consistent, which will now be illustrated in the
particular case of the Homogeneous sine-Gordon theories. 

The HSG theories were constructed in~\cite{HSGClass} at the classical level, 
in~\cite{HSGQuan1} as multiparameter quantum integrable
deformations of conformal field theories, 
and in~\cite{HSGQuan2} as factorised $S$-matrix theories.
Some of their non-perturbative properties have been recently
investigated in~\cite{HSGNp1,HSGNp2,PATRICK,CDF}. 
They are associated with cosets of
the form $G/H$, where
$G$ is a compact simple Lie group of rank
$r$, and $H\simeq U(1)^r$ is a maximal torus of
$G$. In the construction of
section~\ref{Models}, they correspond to triplets
${\go h}\subset {\go g}= {\go f}$, where
${\go g}$ is the Lie algebra of $G$ and ${\go h}$ is the Cartan subalgebra
of ${\go g}$ associated with the maximal torus $H$. 
Moreover, $T_\pm$ are chosen such that the
centraliser of each of them in $g$ is ${\go h}$.
In these theories, it
is easy to understand why not any $T_0\in {\go h}$ leads to a
consistent duality transformation. The reason is that
the set of possible automorphisms $\tau$ in~\rf{ModAction} forms a discrete
group, and only the choices of $T_0$ such that both $\tau$ and
$\tau\cdot\sigma_{T_0}$ are in this group lead to consistent
transformations.

An admissible automorphism $\tau$ has to satisfy two
conditions. The first one is that it leaves the restriction of the
bilinear form
\mbox{$\bil{\>}{\>}$}  to ${\go h}$ invariant, namely
$\bil{\tau(u)}{\tau(v)}= \bil{u}{v}$ for all $u,v\in{\go h}$, to ensure
that the group of gauge transformations~\rf{GaugeT} is anomaly
free~\cite{GWZW}. Since
${\go h}$ is a Cartan subalgebra and $G$ is simple, the restriction of
\mbox{$\bil{\>}{\>}$}  to ${\go h}$ is (proportional to) the Euclidean
metric on ${\fl R}^r$. Therefore, this condition constrains
$\tau$ to be an orthogonal $O(r)$ transformation acting on
${\go h}$~\cite{HSGClass,HSGQuan1}. The second condition arises by
noticing that the gauge transformations~\rf{GaugeT} are not defined
in terms of $\tau$, but in terms of 
$\widehat{\tau}$, which is the lift of
$\tau$ into $H$ defined as follows. Let
$\{h^1\ldots h^r\}$, with
$(h^i)^\dagger=h^i$ and $\bil{h^i}{h^j}=\delta_{ij}$, be a basis of ${\go
h}$, and write a generic element of
$H$ as
$\exp \bigl(2\pi i \vec{\phi}\cdot \vec{h}\bigr)$, where $\vec{\phi}$ is a
$r$-dimensional real vector.  Then,
\begin{equation}
\widehat\tau\left(\exp \bigl(2\pi i \vec{\phi}\cdot \vec{h}\bigr)\right) =
\exp \bigl(2\pi i \tau(\vec{\phi})\cdot \vec{h}\bigr) \>,
\lab{WidehatTau}
\end{equation}
where we have used the same notation for the automorphism $\tau$ acting
on $\vec{\phi}\cdot \vec{h}\in {\go h}$ and for the corresponding linear
transformation acting on
$\vec{\phi}\in {\fl R}^r$; {\em i.e.\/}, $\tau\bigl( \vec{\phi}\cdot
\vec{h}\bigr)\equiv 
\tau(\vec{\phi})\cdot \vec{h}$. This way, the second condition to be
satisfied by
$\tau$ is simply that
$\widehat{\tau}$ is well defined on $H$.\foot{I thank Patrick Dorey for pointing out this condition, which determines that $\tau$ has to
be an element of a discrete group, and was missed in the original papers about
the HSG models.}  In order to
solve it, we have to make the torus structure of $H\subset G$ explicit.
Notice that $\exp
\bigl(2\pi i
\vec{\phi}\cdot
\vec{h}\bigr)$ furnishes a map from ${\fl R}^r$, where $\vec{\phi}$
takes values, onto $H$.  Therefore, $H$ can be identified with ${\fl
R}^r$ factored out by the kernel of this map. This is the set of vectors
$\vec{\phi}\in {\fl R}^r$ mapped onto the unit element of $G$;
{\em i.e.\/}, the vectors that satisfy
\begin{equation}
\exp \Bigl(2\pi i \vec{\phi}\cdot \vec{h}\Bigr) =1\>.
\lab{Monopole}
\end{equation}
This identity has to hold in any representation, and it
is convenient to write it in terms of the weights of $G$. Recall that a
weight $\vec{w}=(w^1\ldots w^r)\in {\fl R}^r$ is the eigenvalue of
$\{h^1\ldots h^r\}$ corresponding to a common eigenvector in some
representation of $G$. The set of these weights is
the `weight lattice' of $G$, which will be denoted
$\Lambda_{\rm w\/}(G)$. Then,~\rf{Monopole} is equivalent to
\begin{equation}
\vec{\phi}\cdot \vec{w} \in {\fl Z}\>,\qquad \forall\> \vec{w}\in
\Lambda_{\rm w\/}(G)\>.
\lab{DualLattice}
\end{equation}
The vectors that satisfy~\rf{DualLattice} span another lattice
$\Lambda_{\rm w\/}^\ast(G)$ known as the `dual lattice' to
$\Lambda_{\rm w\/}(G)$. Consequently, there is a solution
to~\rf{Monopole} for each
$\vec{\phi}\in \Lambda_{\rm w\/}^\ast(G)$.
This provides the identification\foot{
In the rather different context of gauge theories, eq.~\rf{Monopole}
can be recognised as the quantisation condition satisfied by the magnetic
weights of monopoles, which has been solved long ago by Goddard, Nuyts
and Olive in~\cite{OLIVE}, where details about $\Lambda_{\rm
w\/}^\ast(G)$ can be found. $\Lambda_{\rm
w\/}^\ast(G)$ is the weight lattice of the `dual group' to
$G$. In particular, if $G$ is simply connected, in addition to semi-simple,
compact and connected, $\Lambda_{\rm w\/}^\ast(G)$ is the co-root lattice of
$G$, denoted
$\Lambda_{\rm r\/}^\vee (G)$ and defined as the integer
span of the simple co-roots
$\vec{\alpha}_i^\vee=(2/\vec{\alpha}_i^2)\vec{\alpha}_i$, where
$\vec{\alpha}_1 \ldots \vec{\alpha}_r$ form a set of simple roots of
$G$. The group of automorphisms of $\Lambda_{\rm r\/}^\vee (G)$ is the
semidirect product of the Weyl group of $G$ and the group of
automorphisms of the Dynkin diagram of $G$~\cite{HUMP}. In more general cases
where $G$ is not simply connected, 
$\Lambda_{\rm r\/}^\vee (G)$ is always contained in $\Lambda_{\rm
w\/}^\ast(G)$.}
\begin{equation}
H \simeq {{\fl R}^r / \Lambda_{\rm w\/}^\ast(G)}\>.
\lab{Torus}
\end{equation}
Consequently, the requirement that the lift of $\tau$ specified
by~\rf{WidehatTau} is well defined on $H$ constrains $\tau$ to be an
element of the group of automorphisms of $\Lambda_{\rm w}^\ast(G)$,
denoted ${\rm Aut\;}
\Lambda_{\rm w\/}^\ast(G)$, which is a discrete subgroup of $O(r)$.

The conclusion is that the action~\rf{ModAction} specifies a different
HSG model for each choice of $\tau\in {\rm Aut\;} \Lambda_{\rm
w\/}^\ast(G)$ acting on ${\go h}$ according to $\tau\bigl(
\vec{\phi}\cdot
\vec{h}\bigr)\equiv 
\tau(\vec{\phi})\cdot \vec{h}$, and that the models
corresponding to
$\tau$ and
$\tau\cdot \sigma_{T_0}$ are related by a T-duality transformation of the
form~\rf{U1DualAct} for each
$T_0\in{\go h}$ such that $\sigma_{T_0}$ is also in
$\Lambda_{\rm w\/}^\ast(G)$.

An important set of transformations that
leave $\Lambda_{\rm w\/}^\ast(G)$ invariant is provided by the Weyl group
of $G$, denoted ${\cal W}(G)$. It is generated by the reflections
in the hyperplanes orthogonal to the roots of $G$, known as Weyl
reflections, which are the linear transformations
\begin{equation}
w_{\vec{\alpha}}(\vec{\phi}) = \vec{\phi} - 2\> {\vec{\alpha}\cdot
\vec{\phi}\over \vec{\alpha}\cdot \vec{\alpha}}\> \vec{\alpha}
\end{equation}
defined for each root $\vec{\alpha}$ of $G$. Acting on
the Cartan subalgebra, $w_{\vec{\alpha}}$ corresponds to the 
automorphism $\sigma_{T_0}$ defined in~\rf{Reflect} for
$T_0=\vec{\alpha}\cdot\vec{h}$,
\begin{equation}
w_{\vec{\alpha}}(\vec{\phi})
\cdot\vec{h} = \sigma_{\vec{\alpha}\cdot\vec{h}}\bigl(\vec{\phi}\cdot
\vec{h}\bigr)\>.
\end{equation}
Therefore, since $w_{\vec{\alpha}}\in {\rm Aut\;} \Lambda_{\rm
w\/}^\ast(G)$, there is a T-duality transformation of the
form~\rf{U1DualAct} associated with $T_0=\vec{\alpha}\cdot\vec{h}$ for
each root $\vec{\alpha}$, which will be denoted
$D_{T^0}= D_{\vec{\alpha}\cdot\vec{h}}$ using the notation
introduced just before~\rf{Product}.
These transformations can be multiplied according to~\rf{Product},
so that $D_{\vec{\beta}\cdot\vec{h}}\> D_{\vec{\alpha}\cdot\vec{h}}$
relates the HSG models corresponding to the automorphisms $\tau$ and
\mbox{$\tau\cdot (w_{\vec{\alpha}}\cdot w_{\vec{\beta}})$}. This allows
one to associate a T-duality transformation to each Weyl transformation.
Recall that a generic element of ${\cal W}(G)$ is obtained as
the product of a finite number of Weyl reflections, say $\omega =
w_{\vec{\alpha}^{(1)}}\cdot w_{\vec{\alpha}^{(2)}} \cdot \ldots \cdot
w_{\vec{\alpha}^{(n)}}$, where
$\vec{\alpha}^{(1)}\ldots \vec{\alpha}^{(n)}$ are roots of $G$.
Then, the duality transformation associated with $w$ is defined by
\begin{equation}
D_w = D_{\vec{\alpha}^{(n)}\cdot\vec{h}}\> \ldots\>
D_{\vec{\alpha}^{(2)}\cdot\vec{h}}\>
D_{\vec{\alpha}^{(1)}\cdot\vec{h}}\>,
\end{equation}
and it relates the HSG models corresponding to $\tau$ and $\tau\cdot w$.

In general, the semiclassical duality
transformations should not always be expected to correspond to
exact duality symmetries of the quantum theories. However, since the
potential remains invariant in the transformations of the
form~\rf{U1DualAct}, it is natural to expect a correspondence between
exact duality symmetries of the unperturbed $G/U(1)^r$ coset CFT and the
duality symmetries of the quantum HSG theories. The exact duality
symmetries of WZW and coset models have been identified by Kiritsis
in~\cite{KIRplus} by means of the study of their partition
function on the torus. Remarkably, for compact
$G/U(1)^r$ cosets, the exact duality transformations
are in one-to-one relation with the elements of the Weyl group of $G$,
and they correspond to the transformations $D_w$
constructed in the previous paragraph. This leads to conjecture that, for
each
$w\in{\cal W}(G)$,  the semiclassical
T-duality transformation $D_w$ provides an exact
duality symmetry of the $G/U(1)^r$ quantum HSG models.

So far, the duality symmetries constructed in this section have been
presented as examples of transformations of the form~\rf{TDualMass}, which
relate two different dual sigma  models perturbed by the same potential.
Remarkably, the transformations $D_w$ associated with the Weyl transformations
can alternatively be written in the form~\rf{TPoint}, as duality relations
between two different perturbations of the same non-linear sigma model. In
order to make this explicit, let us indicate the dependence of~\rf{ModAction}
on
$T_+$ and $T_-$, the two elements of the Cartan subalgebra ${\go h}$ that
specify the potential,
\begin{equation}
S^{\{\tau\}}[h,A_\pm]\equiv S^{\{\tau\}}_{(T_+,T_-)}[h,A_\pm]\>.
\end{equation}
A Weyl transformation $w\in{\cal W}(G)$ can always be lifted
to an inner automorphism of $G$, which ensures the existence of a (non
unique) constant group element
$\gamma_w\in G$ such that
$w(u) = \gamma_w^{-1} \>u \>\gamma_w$ for all $u\in{\go h}$. This leads to the
identity
\begin{equation}
S^{\{\tau\cdot w\}}_{(T_+,T_-)}[h,A_\pm] =
S^{\{\tau\}}_{(T_+,w(T_-))}[\gamma_w^{-1} h,w(A_\pm)]
\>.
\lab{FieldTrans}
\end{equation}
Therefore, up to a change of the field variables, the duality
transformation $D_w$ becomes
\begin{equation}
S^{\{\tau\}}_{(T_+,T_-)} \buildrel D_w \over{\hbox to
45 pt{\rightarrowfill}} S^{\{\tau\cdot w\}}_{(T_+,T_-)} 
\Buildrel {h \>\rightarrow \>\gamma_w h} \over{\hbox to
90pt{\rightarrowfill}} \under{A_\pm\> \rightarrow\> w^{-1}(A_\pm)}
S^{\{\tau\}}_{(T_+,w(T_-))}\>,
\lab{DualAct}
\end{equation}
which is a duality relation of the form~\rf{TPoint} where the potential
transforms according to
\begin{equation}
{\cal U}\equiv \lambda\> \bil{T_+}{h^{-1}\>T_- h}
\;\; {\hbox to
30pt{\rightarrowfill}}\;\; 
\widetilde{{\cal U}} \equiv \lambda\> \bil{T_+}{h^{-1}\> w(T_-) h}
\>.
\lab{PotTransf}
\end{equation}

For the Complex sine-Gordon theory, which is recovered for $G=SU(2)$, the only
non-trivial Weyl transformation is $w=-I$. Then,~\rf{PotTransf}
becomes simply
${\cal U}\rightarrow
\widetilde{{\cal U}}=-\>{\cal U}$. This relates the two phases
of the model, which are characterised by the sign of its unique coupling
constant. This case will be analysed in more detail in the next section. 

\subsubsection{T-duality and the phases of
the HSG models.}
\label{HSGPhase}

~\indent
For general HSG models, the transformations~\rf{PotTransf} also 
relate the different phases of the model, which are in
one-to-one relation with the elements of the Weyl group of $G$. This can be
proved as follows.

First, we have to characterise the phases of the HSG models. This can be
done by studying the form of the manifold of vacuum field
configurations, which correspond to the minima of the the
potential~\rf{Potential}. A constant field configuration
$h_0$ is a minimum of~\rf{Potential} if it
satisfies two conditions~\cite{HSGClass}. The first one is
\begin{equation}
\bigl[T_+, h_0^{-1} T_- h_0]=0\>,
\lab{CondPhaseA}
\end{equation}
which ensures that the potential is stationary at $h_0$. 
In the HSG models, the
centralisers of $T_+$ and $T_-$ coincide with the
Cartan subalgebra ${\go h}$. Then,~\rf{CondPhaseA} implies that the inner
automorphism of $g$ generated by
$h_0$ leaves ${\go h}$ fixed and, thus, it corresponds to a Weyl
tranformation of
$G$. Therefore, the solutions to~\rf{CondPhaseA} are of the form
$h_0=\gamma_\sigma$, where $ \gamma_\sigma^{-1} \>u \>\gamma_\sigma=
\sigma(u)$ for all
$u\in{\go h}$, and $\sigma\in{\cal W}(G)$. 

The second condition is needed to
ensure that $h_0=\gamma_\sigma$ actually corresponds to a minimum
of the potential, which is equivalent to require that all the small
fluctuations around
$h_0$ have real non-vanishing masses. Let $T_\pm=\pm i\lav_\pm \cdot \vec{h}$,
and write $\gamma_\sigma^{-1} \>T_-
\>\gamma_\sigma \equiv -i \sigma(\lav_-)\cdot \vec{h}$ using the conventions
introduced just after~\rf{WidehatTau}. Then, the mass spectrum of
the small fluctuations around $h_0=\gamma_\sigma$ is given by~\cite{HSGClass}
\begin{equation}
m_{\alv}^2 = 4\pi\lambda \bigl(\alv\cdot \lav_+\bigr) \bigl(\alv\cdot
\sigma(\lav_-)\bigr)
\end{equation}
for each root $\alv$ of $g$. Thus, the condition that $h_0=\gamma_\sigma$
corresponds to a minimum is that all these numbers are
strictly positive. If $\lambda>0$, this requires that $\lav_+$ and
$\sigma(\lav_-)$ are inside the same Weyl chamber of ${\go h}$. 
Recall that the set of hyperplanes orthogonal to all the roots of $G$
partitions the Euclidean space ${\fl R}^r$ into disjoint connected components
called Weyl chambers. The Weyl group of $G$ permutes the Weyl chambers, so
that each two chambers are related by a Weyl transformation. Once a system of
simple roots $\Delta=\{\vec{\alpha}_1 \ldots
\vec{\alpha}_r\}$ is chosen, there is one Weyl chamber, denoted $C(\Lambda)$,
such that any $\vec\phi\in C(\Lambda)$ satisfies $\alv_i\cdot \vec{\phi}>0$ for all
$i=1\ldots r$. $C(\Lambda)$ is called the `fundamental Weyl
chamber'~\cite{HUMP}. We will choose the system of simple roots such
that $\lav_+\in C(\Delta)$.

We will now argue that there is a different phase for each $\sigma\in{\cal
W}(G)$ that is specified by the domain where
$\lav_-$ takes values. In particular, the $\sigma$-phase 
corresponds to
\begin{equation}
\lav_-\in\sigma^{-1}\bigl(C(\Delta)\bigr)\>,
\lab{PhaseDef}
\end{equation}
so that that $\lav_-$ takes
values in disjoint components of ${\fl R}^r$ for different phases. This will
be supported by showing that the form of the manifold of
vacuum field configurations depends on $\sigma$. By construction, the 
vacuum configurations in the 
$\sigma$-phase are of the form
$h_0=\gamma_\sigma$, where $\gamma_\sigma$ is specified by the condition
$\gamma_\sigma^{-1} \>u \>\gamma_\sigma= \sigma(u)$ for all $u\in{\go h}$.
Its general solution can be written as
\begin{equation}
\gamma_\sigma = \widehat{\gamma}_\sigma \>\ex{v}\>,
\lab{VacA}
\end{equation}
where $\widehat{\gamma}_\sigma$ is a fixed particular solution and $v$ is any
element of ${\go h}$. This shows that the space of field
configurations of this type is isomorphic to $U(1)^r$. However, not all these
configurations are physical. Some of them become identified
under the action of the $\tau$-dependent group of gauge
transformations~\rf{GaugeT}. In particular,
\begin{equation}
\widehat{\gamma}_\sigma \ra \ex{v}\> \widehat{\gamma}_\sigma\> \ex{-\tau(v)} =
\widehat{\gamma}_\sigma \bigl( \widehat{\gamma}_\sigma^{-1}\> \ex{v}\>
\widehat{\gamma}_\sigma\bigr)\>
\ex{-\tau(v)} =
\widehat{\gamma}_\sigma\> \ex{(\sigma-\tau)(v)}\>,
\lab{VacAA}
\end{equation}
which is not trivial for each $v\in{\go h}$ such that
$(\sigma-\tau)(v)\not=0$. The rank of the linear transformation
$\sigma-\tau$ is the number of linear independent `$v$' which are not in its
kernel. Therefore, the set of field configurations of the form~\rf{VacA} that
become identified with
$\widehat{\gamma}_\sigma$ under the group of gauge transformations is
isomorphic to $U(1)^{{\rm rank\/}(\sigma-\tau)}$. This proves that the
manifold of physical vacuum configurations is
\begin{equation}
\{h_0 \} \simeq U(1)^{r-{\rm rank\/}(\sigma-\tau)}\>,
\lab{Vacb}
\end{equation}
which does depend on $\sigma$, and justifies the proposed identification of
the phases of the HSG models.

The comparison of~\rf{PotTransf} and~\rf{PhaseDef} confirms that, for each
$w\in{\cal W}(G)$, the duality transformation $D_w$ specified by~\rf{DualAct} 
indeed provides a relationship between two different phases of the model;
namely, it relates the $\sigma$-phase with the $\sigma\cdot w^{-1}$-phase.

Notice that~\rf{Vacb} provides a condition for
the existence of a phase where the vacuum of the model is
not degenerate. It requires that ${\rm rank\/}(\sigma-\tau)=r$ for
some $\sigma\in{\cal W}(G)$. This 
clarifies the meaning of the condition deduced in~\cite{HSGClass} to ensure
that the potential has no flat directions and, hence, that the theory has a
mass gap. As an example where this condition is met, consider the
HSG theories where the group of gauge transformations is of axial type, which
corresponds to $\tau=-I$. Then, the condition is trivially satisfied for
$\sigma=I$. In contrast, if
$-I\in{\cal W}(G)$, then ${\rm rank\/}(\sigma-\tau)=0$ for $\sigma=-I$, and
the vacuum of the theory is maximally degenerate in the  corresponding phase.

\sect{T-duality in the CSG theory.} 
\label{CSG}

~\indent
As a prototypical example, we now discuss in detail one of the simplest
integrable theories that exhibits a duality symmetry of the form~\rf{TPoint}:
the Complex sine-Gordon (CSG) model. It has two different phases
corresponding  to the two signs of its unique coupling constant, and they turn
out to be related by T-duality. This duality symmetry was already pointed out
by Bakas~\cite{BAKAS}, and an explicit transformation rule was constructed by
Park and Shin~\cite{CSGDUALITY}. Nevertheless, the latter is only valid
on-shell. In contrast, we will  show that that the
two phases are related off-shell by a canonical transformation between
the phase spaces.  The proper understanding of the T-duality symmetry helps to
clarify the nature of the already known CSG soliton solutions. In one of the
phases, due to the trivial vacuum structure, the soliton solutions are
topologically trivial -- they are of the form $[0,Q^{\rm N}]$. Then, the
duality map provides a topological interpretation for them in the
other phase. This leads to the discovery of Bogomol'nyi-like 
bounds for the energy saturated by the usual one-soliton solutions in
both phases which, to our knowledge, have been overlooked in the
literature.

\subsection{Basics of the CSG model.}

~\indent
The CSG model is defined by the
relativistic two-dimensional Lagrangian
\begin{equation}
{\cal L}_{\rm CSG}= {1\over 4\pi \beta^2}\left({\partial_\mu
\psi\>\partial^\mu \psi^\ast\over 1- \psi \psi^\ast} - \lambda\>
\psi \psi^\ast\right)\>,
\lab{CSGLag}
\end{equation}
where $\psi=\psi(t,x)$ is a complex scalar field, and $\lambda$ and
$\beta$ are real coupling constants; $\lambda$ is dimensionful, while
$\beta$ is dimensionless and plays no role in the classical theory. 
The Lagrangian is invariant under the global $U(1)$ transformations
$\psi(t,x)\rightarrow \ex{i\alpha}\psi(t,x)$. A more convenient form of the
CSG Lagrangian is obtained if we substitute
$\psi=\sin\eta\> \ex{i\phi}$, with $\eta$ and $\phi$ real fields.
Then,~\rf{CSGLag} becomes
\begin{equation}
{\cal L}_{\rm CSG}= {1\over 4\pi \beta^2}\left(\partial_\mu
\eta\>\partial^\mu \eta + \tan^2\eta\>\partial_\mu
\phi\>\partial^\mu \phi  - \lambda\>
\sin^2\eta\right)\equiv {\cal L}_{\rm
CSG}(\phi,\eta; \lambda)\>,
\lab{CSGLagTwo}
\end{equation}
which allows one to make explicit the relationship
with the usual sine-Gordon model: it is recovered just by taking the field
$\phi$ to be constant. Notice that both~\rf{CSGLag} and~\rf{CSGLagTwo} are
Lagrangians of the form~\rf{SigmaModPot}.

The CSG model was originally introduced by Lund and Regge
to describe relativistic vortices in a superfluid~\cite{LUND} and,
independently, by Pohlmeyer as a reduction of the $O(4)$ non-linear sigma
model~\cite{POHL}. The geometric interpretation of the model has
been discussed by Bakas~\cite{BAKAS}. It depends crucially on the sign of
$\lambda$, which manifests that the model has two different phases:
one for $\lambda>0$ and another one for $\lambda<0$. 
More recently, the classical aspects of the CSG model in the presence of a
boundary have been addressed by Bowcock and Tzamtzis~\cite{BT}.

At the quantum level, the CSG model has been investigated using both
ordinary perturbation theory~\cite{dVM} and semiclassical
techniques~\cite{dVMSemi,DH}. However, the way to properly define the
theory non-perturbatively follows from the work of Bakas, who showed at the
classical level that the CSG model can be formulated in terms of a gauged
WZW action~\cite{BAKAS} (see also~\cite{PARK}) so that the model is
defined by an action of the form~\rf{ModAction} associated with the
coset $SU(2)/U(1)$. In this case, it is convenient to make the bosonic
field $h$ take values in the fundamental representation of $SU(2)$, and to
choose the bilinear form as $\bil{u}{v}={\rm Tr\/}(u v)$. Then, if we
call $\sigma_1,\sigma_2,\sigma_3$ the usual
Pauli matrices, the generator of the $U(1)$ factor can be chosen to be
$T^0={i\over\sqrt{2}}\> \sigma_3$, and the potential becomes
\begin{equation}
U(h)= +{\lambda\over 16\pi} \> {\rm Tr\/}\left(h\sigma_3 h^{-1}\sigma_3
\right)\>.
\lab{PotentialCSG}
\end{equation}
Concerning the automorphism $\tau$, as explained in section~\ref{HSG} it has
to belong to the group of automorphisms of the dual to the
weight lattice of $SU(2)$, which coincides with the root lattice of $SU(2)$.
This group is isomorphic to the cyclic group ${\fl Z}_2$ and, thus, there are
only two choices. It can be either
$\tau=+I$ or $\tau=-I$, which lead to $U(1)$ gauge transformations of
vector or axial form, respectively. For this reason, it is
convenient to introduce the notation
\begin{equation}
S_{\rm CSG}^{\{+I\}}\equiv S_{\rm CSG}^{\rm V}\>, \qquad 
S_{\rm CSG}^{\{-I\}}\equiv S_{\rm CSG}^{\rm A}\>.
\lab{AVAct}
\end{equation}

The connection between $S_{\rm CSG}^{\{\tau\}}$ and the CSG
Lagrangian ${\cal L}_{\rm CSG}$ is recovered as follows. Consider the
following parametrisation of the $SU(2)$ field $h$,
\begin{equation}
h= 
\begin{pmatrix}
u& -i v^\ast \\
-i v& u^\ast\\
\end{pmatrix}\>, \quad {\rm with}\quad |u|^2 + |v|^2 =1\>.
\lab{SU2}
\end{equation}
Then, integrating out the non-dynamical fields $A_\pm$ by means of their
equations of motion, the two actions provided by~\rf{ModAction}
become
\begin{eqnarray}
&&S_{\rm CSG}^{\rm V}[h,A_\pm;\lambda] = {k\over 4\pi}
\int d^2 x \left({\partial_\mu u\>\partial^\mu u^\ast\over 1- u u^\ast} -
\lambda\> (uu^\ast-{1\over2})\right)\nn[6pt]
&&\qquad{\rm and}\qquad S_{\rm CSG}^{\rm A}[h,A_\pm;\lambda] = {k\over
4\pi}
\int d^2 x \left({\partial_\mu v\>\partial^\mu v^\ast\over 1- v v^\ast} +
\lambda\> (vv^\ast-{1\over2})\right)\>.
\lab{CSGWZW}
\end{eqnarray}
Up to a constant shift of the energy density, both of them
correspond to the Lagrangian of the CSG model given by~\rf{CSGLag} if
we identify the coupling constant $1/\beta^2$  with $k$, the level of the
$SU(2)_k/U(1)$ coset CFT, and leave the sign of $\lambda$ free. It is
worth noticing that, in this approach, the singularity of~\rf{CSGLag} at
$|\psi|=1$ comes from the elimination of
$A_\pm$, but it does not correspond to a real singularity of the
action~\rf{ModAction}.

At the quantum level, the action~\rf{ModAction} for the coset
$SU(2)/U(1)$ provides a Lagrangian description of the the theory
of ${\fl Z}_k$ parafermions -- the
$SU(2)_k/U(1)$ coset conformal field theory -- perturbed by their first
thermal operator, which is known to be quantum integrable~\cite{PINT}. To
be specific,~\rf{ModAction} is an action of the form~\rf{PCFT}
where, in this case, $S_{\rm CFT}$ denotes an action for the $SU(2)_k/U(1)$
coset conformal field theory, with central charge $2(k-1)/(k+2)$, and
the perturbation is defined by the spinless primary field corresponding to
the first thermal operator, whose conformal dimension is $\Delta= 
2/(k+2)$.

It is important to stress that~\rf{CSGWZW} gives a complete
description of the perturbed gWZW model~\rf{ModAction} only in the
large~$k$ limit, which corresponds to both the semiclassical and
weak coupling limits. Therefore, the CSG model provides an explicit
Lagrangian formulation of the theory of
${\fl Z}_k$ parafermions perturbed by the first thermal operator when $k$
is large, provided that the CSG coupling constant $\beta^2$ is identified
with $1/k$, which in this context is
also required to make sense of the WZW action in~\rf{gWZW}. An independent
motivation for the quantisation of the CSG coupling constant is provided
by the semiclassical analysis of the CSG scattering amplitudes, where it
appears as a condition to ensure that the theory admits a factorisable
$S$-matrix~\cite{DH}. 

In turn, the perturbed 
${\fl Z}_k$ parafermionic theory can be used as a non-perturbative
definition of the CSG theory beyond the large $k$ limit.
This integrable perturbed CFT develops a finite correlation length and is
described by the minimal factorised $S$-matrix associated with the Lie
algebra $a_{k-1}$. This is independent of the sign of its coupling
constant~\cite{PINT}, which confirms that the T-duality identification
between the two phases of the model persists in the quantum theory, where
it is a consequence of the order-disorder
duality symmetry of the unperturbed ${\fl Z}_k$ parafermionic
theory~\cite{PARAF,BAKAS} (see also~\cite{CABRA}).

\subsection{Off-shell T-duality in the CSG model.}

~\indent
In this case the non-dynamical gauge fields have only one component,
$A_\pm= a_\pm T^0$, and the automorphism defined in~\rf{Reflect} is
$\sigma_{T_0}=-I$. Therefore, the duality transformation~\rf{U1Dual}
associated with the $U(1)$ global symmetry corresponding to
$\psi(t,x)\rightarrow
\ex{i\alpha}\psi(t,x)$ reads
\begin{equation}
\left(S_{\rm CSG}^{\rm V}\right)^{\rm D}\bigl[\ex{\beta\>T^0}\> h_0\>
\ex{\gamma\>T^0},A_+,A_-;
\lambda\bigr] = S_{\rm CSG}^{\rm A}\bigl[\ex{\beta\>T^0}\> h_0\>
\ex{-\gamma\>T^0},A_+,A_-^{'}; \lambda\bigr]\>,
\lab{CSGDual}
\end{equation}
where $A_-^{'}= -A_- + 2\partial_- \gamma\> T^0$, which, in terms of the
Lagrangian~\rf{CSGLagTwo}, is equivalent to\foot{The field $\phi$ introduced
in~\rf{CSGLagTwo} and used along section~\ref{CSG} is related to the
parametrisation~\rf{WZWfield} by means of $\phi=(\beta+\gamma)/\sqrt{2}$.
The canonical
transformation that generates~\rf{CSGDual} reads $\Pi_\phi= (k/2\pi)
\partial_x \phi^{\rm D}$ and $\Pi_{\phi^{\rm D}}= (k/2\pi)
\partial_x \phi$, to be compared with~\rf{CanTransf}.\label{footNorm}}
\begin{equation}
{\cal L}_{\rm CSG}(\phi,\eta;\lambda) \rightarrow {\cal L}_{\rm
CSG}^{\rm D}(\phi,\eta;\lambda)= {k\over 4\pi
}\left(\partial_\mu
\eta\>\partial^\mu \eta +  \cot^2\eta\>\partial_\mu
\phi\>\partial^\mu \phi  - \lambda\>
\sin^2\eta\right)\>,
\lab{CSGLagDual}
\end{equation}
and we have already substituted the coupling constant $\beta^2$ by $1/k$.
Notice that the potential $U=(k/4\pi)\> \lambda\>
\sin^2\eta$ does not change, which
confirms that the duality transformation is of the form~\rf{TDualMass}.

The dual Lagrangian~\rf{CSGLagDual} is related to the original
one~\rf{CSGLagTwo} by means of the trivial field transformation
$\widetilde{\eta}={\pi\over2}-\eta$ as follows
\begin{equation}
{\cal L}_{\rm CSG}^{\rm D}(\phi,\eta;\lambda) = {\cal L}_{\rm
CSG}(\phi,\widetilde\eta;\> -\lambda) - {k\over 4\pi}\>\lambda\> .
\lab{LagPointTrans}
\end{equation}
Therefore, up to the constant shift of the energy density, the effect of the
T-duality transformation is just to change the sign of the coupling
constant $\lambda$ and, hence, to provide an off-shell relationship
between the two phases of the model. The resulting transformation is of the
form~\rf{TPoint} with $\widetilde{{\cal U}}=-\>{\cal U}$.

In the gauged WZW formulation, the transformation~\rf{LagPointTrans}
corresponds to~\rf{FieldTrans} that, in this case, reads
\begin{equation}
S_{\rm CSG}^{\rm V}\bigl[h,A_+,A_-;\lambda\bigr] = S_{\rm CSG}^{\rm
A}\bigl[i\sigma_1 h,-A_+,-A_-;-\lambda\bigr] \>.
\lab{OrderDisorder}
\end{equation}
Actually, it was already pointed out by Bakas~\cite{BAKAS} that this is a
manifestation of the symmetry under the order-disorder (Kramers-Wannier)
duality transformation of the theory of ${\fl Z}_k$ parafermions, in the
perturbed conformal field theory language. Indeed, according to~\rf{CSGWZW},
$S_{\rm CSG}^{\rm V}$ provides a description of the theory in terms of
$u$, which is the gauge invariant component of the WZW field $h$ in~\rf{SU2}
with respect to vector gauge transformations, while the elementary field in
$S_{\rm CSG}^{\rm A}$ is $v$, the gauge invariant component with respect
to axial gauge transformations. In the context of the theory
of parafermions~\cite{PARAF,BAKAS}, $u$ and $v$ represent the
spin, $\hat\sigma_1=\phi^{({1})}_{1,1}$, and dual-spin, 
$\hat\mu_1=\phi^{({1})}_{1,-1}$, fields, respectively. They are the
diagonal and off-diagonal components of the WZW field in the fundamental
representation of $SU(2)$, whose conformal dimension is
$\Delta=\overline\Delta=(k-1)/2k(k+2)$. 
Similarly, the composite field ${\rm Tr\/}\left(h\sigma_3 h^{-1}\sigma_3
\right)=2(u u^\ast -v v^\ast)$ represents the first thermal operator
$\phi^{(2)}_{0,0}$, whose conformal dimension is $2/(k+2)$. This way, the
duality transformation~\rf{OrderDisorder} corresponds simply to
$u=\hat\sigma_1\leftrightarrow
\hat\mu_1=v$.

\subsection{On-shell T-Duality and CSG soliton solutions.}
\label{CSGSolitons}

~\indent
As explained in section~\ref{OnShell}, on shell, 
eqs.~\rf{CanTransf} and~\rf{Pseudo} provide a
map between the solutions to the equations of motion of~\rf{CSGLag},
\begin{equation}
\partial_\mu\partial^\mu \psi \>+\>\psi^\ast\> {\partial_\mu\psi
\>\partial^\mu\psi\over 1-\psi\psi^\ast} \>+\>\lambda\>
\psi(1-\psi\psi^\ast)=0\>,
\lab{EOM}
\end{equation}
in the two phases of the model. Indeed, if we denote a generic solution
of~\rf{EOM} by $\psi^{(\lambda)}$, the duality transformation
$(\phi,\eta;\lambda) \rightarrow (\phi^{\rm D}, {\pi\over2}-\eta;-\lambda)$
corresponds to
\begin{equation}
\psi^{(\lambda)} = \sin\eta\> \ex{i\phi} \longrightarrow
\psi^{(-\lambda)} = \cos\eta\> \ex{i\phi^{\rm D}}, 
\lab{OnShellDual}
\end{equation}
where $\phi^{\rm D}$ is the solution to the equations of the
canonical transformation understood as partial differential equations for
$\phi^{\rm D}$,
\begin{equation}
\partial_x \phi^{\rm D} = +\tan^2\eta\> \partial_t \phi\>, \qquad
\partial_t \phi^{\rm D} = +\tan^2\eta\> \partial_x \phi\>.
\lab{OnShellCan}
\end{equation}
This on-shell duality, or pseudoduality, transformation coincides with
the transformation constructed by Park and Shin within the gauged WZW
formulation of the CSG model using a particular choice of the gauge fixing
prescription~\cite{CSGDUALITY}. There,~\rf{OnShellDual} reads simply
$h\rightarrow i\sigma_1 h$, where $h$ is a solution to the equations
of motion of~\rf{ModAction} with the gauge fixed by the
condition $A_\pm=0$. Using the parametrisation~\rf{SU2}, it is equivalent
to interchange the roles of $u$ and $v$ and simultaneously change the sign
of $\lambda$.\foot{The equations~\rf{OnShellCan} were
originally written in~\cite[section~7]{DUALOLD} as transformations that change
the boundary conditions satisfied by the soliton solutions, without making
any reference to duality.}

The $U(1)$ global symmetry of the CSG
Lagrangian~\rf{CSGLag} leads to the conserved
Noether current
\begin{equation}
J_\mu^{\rm N} = {k\over 4\pi}\>i\> {\psi\> \partial_\mu \psi^\ast
-\psi^\ast\>\partial_\mu
\psi\over 1-\psi\psi^\ast} = {k\over 2\pi}\> \tan^2\eta\>
\partial_\mu
\phi \>.
\lab{NoetherCSG}
\end{equation}
As explained in section~\ref{OnShell}, there is also a topological current
associated with the field $\phi$, which, in this case, is convenient to
define as
\begin{equation}
J_\mu^{\rm T} = -{1\over 2}\> \varepsilon_{\mu\nu}\>\partial^\nu
\phi\>, \qquad \varepsilon_{01}=+1\>. 
\end{equation}
Then, the eqs.~\rf{OnShellCan} can simply be written as 
\begin{equation}
J^{\rm (D)\>T}_\mu = {\pi\over k}\> J^{\rm N}_\mu\>, \qquad
J^{\rm T}_\mu = {\pi\over k}\> J^{\rm (D)\> N}_\mu\>.
\lab{NoetTop}
\end{equation}

Let us briefly review the main features of the already known soliton
solutions of the CSG model from the perspective of
eqs.~\rf{NoetTop}.\footnote{We will only consider relativistic solitons
defined in 1+1 Minkowski space. Solutions of the CSG equation in Euclidean
0+2 space have been constructed in~\cite{EUCLIDEAN}.} 
First, consider the
phase
$\lambda>0$, where the coupling constant $\lambda$ can be properly
understood as a squared-mass parameter. Here, the potential of the CSG
model has a unique minimum at
$\psi=0$, and the boundary conditions to be satisfied by the soliton
solutions are
\begin{equation}
\psi(t,x) 
\buildrel {x\rightarrow \pm\infty}\over{\hbox to
45pt{\rightarrowfill}} 0\>.
\lab{ElecBC}
\end{equation}
Therefore, in this phase, they
carry a non-trivial $U(1)$ Noether charge but no
topological charge. 
One-soliton solutions of this kind
were originally constructed by Getmanov~\cite{GETMANOV}. In their rest
frame, they are given by periodic time-dependent field configurations
rotating in the internal
$U(1)$ space of the form
\begin{equation}
\psi(t,x)=  {\cos\alpha \over \cosh( \sqrt{\lambda}\>\cos\alpha\>
x) }\> \exp(i\sqrt{\lambda}\> \sin\alpha\> t)\>,
\lab{ElectricSol}
\end{equation}
and there is a different one for each value of
the real parameter $\alpha\in(-\pi/2,+\pi/2)$. The classical $U(1)$ Noether
charge and mass carried by this solution are given by
\begin{equation}
Q^{\rm N}= {k\over \pi}\> \left({\rm
sign\/}[\alpha]\>{\pi\over2}-\alpha
\right), \qquad M= {k\over \pi}\> \sqrt{\lambda}  \cos\alpha 
= {k\over \pi}\>\sqrt{\lambda}\>  \bigl|\sin\left(\pi Q^{\rm N}/
k\right)\bigr|\>.
\lab{ChargeMass}
\end{equation}
Using the notation introduced in section~\ref{OnShell},
these solitons are labelled by $[\omega,Q^{\rm N}]\equiv [0,Q^{\rm
N}]^{(+)}$, where the superscript~$^{(+)}$ indicates that they are
solutions to the equations of motion for $\lambda>0$.

In contrast, since $|\psi|^2\leq1$, 
the minima of the potential in the phase $\lambda<0$ correspond to
\hbox{$|\psi|=1$}, and there is an infinite number of them related to each
other by the global $U(1)$ symmetry. In their rest frame, the soliton
solutions interpolating between these minima are
time-independent field configurations that satisfy boundary conditions of
the form
\begin{equation}
\psi(t,x) 
\buildrel {x\rightarrow \pm\infty}\over{\hbox to
45pt{\rightarrowfill}} \ex{i \phi_\pm}
\lab{TopBC}
\end{equation}
and, thus, carry the topological charge
\begin{equation}
Q^{\rm T}=\int_{-\infty}^{+\infty} dx\> J_0^{\rm
T}={\phi_+-\phi_-\over2}\>.
\lab{SolTopCharge}
\end{equation}
Solutions of this kind
were constructed by Lund and
Regge~\cite{LUND},
\begin{equation}
\psi(t,x)=
i\> \ex{i\> {\phi_+ +\phi_-\over2}}\> \left(\sin Q^{\rm T} 
\tanh\bigl(\sqrt{-\lambda}\>
\bigl|\sin Q^{\rm T}\bigr|\> x\bigr) -i\cos Q^{\rm T}\right)\>.
\lab{LundSol}
\end{equation}
Their mass is given by
\begin{equation} 
M(Q^{\rm T})={k\over \pi}\> \sqrt{|\lambda|}\;\; \bigl|\sin Q^{\rm
T} \bigr|\>
\lab{MassForm}
\end{equation}
and their $U(1)$ Noether charge vanishes. Using the notation
introduced in section~\ref{OnShell}, these solitons are labelled by
$[\omega,Q^{\rm N}]\equiv [2Q^{\rm T},0]^{(-)}$, where~$^{(-)}$ indicates that
they are solutions for $\lambda<0$.

T-duality relates the soliton solutions in both
phases. Namely, if $(k/\pi)\> Q^{\rm T}$ equals the Noether charge carried
by the solution~\rf{ElectricSol}, then~\rf{LundSol} is
mapped into~\rf{ElectricSol} under the pseudoduality
transformation~\rf{OnShellDual},
\begin{equation}
[0,Q^{\rm N}]^{(+)} \buildrel {\rm T-duality}\over{\hbox to
60pt{\rightarrowfill}} [(2\pi/k)\> Q^{\rm N},0]^{(-)}\>.
\lab{TdualSol}
\end{equation}
As explained in section~\ref{OnShell}, the
fact that $\phi$ is a compact field implies, through the duality
transformation, that the value of the Noether charge $Q^{\rm N}$ is defined
modulo some period characterised by $k$. In our case, $\phi\sim \phi +
2\pi{\fl Z}$ translates into $Q^{\rm N}\sim Q^{\rm N} + k{\fl Z}$
which, in particular, resolves the apparent discontinuous dependence of
$Q^{\rm N}$ on the value of $\alpha$ in~\rf{ChargeMass}. Correspondingly,
$Q^{\rm T}\sim Q^{\rm T} + \pi{\fl Z}$.

In the $\lambda>0$ phase, solitons are obviously not topological in nature,
and their Noether charge $Q^{\rm N}$ can take any real value, which makes
their stability unclear~\cite{CSGDUALITY,BT}. In the other
phase,
$\lambda<0$, the situation is similar because, as explained in
section~\ref{OnShell}, the conservation of
$Q^{\rm T}$ does not rely on topological arguments and this charge can also
take any real value. The classical stability of the CSG solitons will be
clarified in the next section by showing that they saturate Bogomol'nyi-like
bounds for the energy.

In the quantum theory, the situation is different. Under
quantisation, the Noether charge carried by the solitons in the
$\lambda>0$ phase becomes quantised. The precise form of this
quantisation was found in~\cite{dVMSemi,DH} by applying the Bohr-Sommerferld
(BS) quantisation rule to the periodic soliton  configurations provided
by~\rf{ElectricSol}; {\em i.e.\/}, $S+MT
=2\pi n$ where $n$ is a positive integer, $T$ is the period of the soliton
solution, $M$ is its mass, and $S$ is its action.
For the CSG model, $S+MT= 2\pi\> Q^{\rm N}$, and the BS rule implies that
$Q^{\rm N}$ has to be integer. However, we have already shown that $Q^{\rm N}$
is only defined modulo~$k$, namely $Q^{\rm N}\sim Q^{\rm N} + k{\fl Z}$. This
is consistent with the built-in ambiguity in the definition of the WZW action
in~\rf{ModAction} and, hence, in the combination $S+MT$, which is  precisely
of the form $2\pi k\> {\fl Z}$ with our normalisations. Therefore, the
semiclassical quantisation of the   solitons in the $\lambda>0$ phase
provides exactly $k-1$ non-topological {\em stable\/} solitons with $U(1)$
Noether (electric) charges
\begin{equation}
Q^{\rm N} = n \>, \qquad n =1\ldots k-1 {\rm \;\;
mod\;\;} k\>.
\lab{ElSpec}
\end{equation}
Correspondingly, via the T-duality transformation, the
quantisation of the Noether charge carried by the solitons in the
$\lambda>0$ phase implies the quantisation of the topological (magnetic)
charge carried by the dual solitons. Taking~\rf{TdualSol} into account, the
resulting allowed values of the topological charge are
\begin{equation}
Q^{\rm T} = {\pi \over k} \> n \>, \qquad n
=1\ldots k-1 {\rm
\;\; mod\;\;} k\>.
\lab{MagSpec}
\end{equation}
In other words, and using~\rf{TopBC}, the CSG fields corresponding to the
quantum soliton solutions in the $\lambda<0$ phase satisfy the
parafermionic boundary conditions
\begin{equation}
\psi(t,+\infty)= \exp\Bigl(i\> {2\pi\over k}\>n\Bigr)\> \psi(t,- \infty)\>,
\end{equation}
which is in agreement with the well known breaking of the classical global
$U(1)$ symmetry into a discrete ${\fl Z}_k$ symmetry after
quantisation~\cite{ParafCos}. It is worth noticing that, in both phases, the
resulting quantum spectrum~\rf{ElSpec} and~\rf{MagSpec}, and the the
corresponding mass formulae~\rf{ChargeMass} and~\rf{MassForm}, match the
spectrum of stable particles of the minimal factorised $S$-matrix theory
associated with the Lie algebra $a_{k-1}$, which describes the integrable
theory of ${\fl Z}_k$ parafermions perturbed by the first thermal
operator~\cite{PINT}.

\subsection{Bogomol'nyi-like bounds in the CSG model.}
\label{BPSBounds}

~\indent
The results of section~\ref{OnShell} suggests a topological interpretation
of the one-soliton solutions in the $\lambda<0$ phase, in the sense that they
are characterised by the boundary values of the CSG field.
This interpretation is supported by the fact that they actually saturate a
Bogomol'nyi-like bound, which can be deduced as follows. The energy density
corresponding to a time independent field configuration
$\psi=\psi(x)$ is
\begin{equation}
{\cal H}_{\rm CSG} = {k\over 4\pi }\left({\partial_x
\psi\>\partial_x \psi^\ast
\over 1- \psi \psi^\ast} - \lambda\>
(1-\psi \psi^\ast)\right) +{k\over 4\pi
}\> \lambda\>,
\lab{CSGHam}
\end{equation}
which can be written as
\begin{equation}
{\cal H}_{\rm CSG}= {k\over 4\pi}\left({|\partial_x
\psi\> -i\ex{i\Omega} \sqrt{-\lambda}\> (1- \psi \psi^\ast)|^2
\over 1- \psi \psi^\ast} 
+ 2\sqrt{-\lambda}\>
\partial_x{\rm Im\/}(\ex{-i\Omega}\psi)\right)
+{k\over 4\pi
}\> \lambda\>,
\lab{CSGHamB}
\end{equation}
where $\Omega$ is an arbitrary constant real number.
Taking the boundary conditions~\rf{TopBC} into account, and recalling that
$|\psi|^2\leq1$, eq.~\rf{CSGHamB} leads to
\begin{equation} 
M^{(-)}=\int_{-\infty}^{+\infty} dx\> \left({\cal H}_{\rm CSG} -
{k\over 4\pi }\> \lambda \right)\geq
{k\over 2\pi}\> \sqrt{-\lambda}\;\; \Bigl(\sin(\phi_+-\Omega) -
\sin(\phi_- -\Omega)\Bigr)\>,
\lab{BPSA}
\end{equation}
where the superscript~$^{(-)}$ indicates that $\lambda<0$.  
The most stringent bound is achieved by choosing
$\ex{i\Omega}= {\rm sign\/}\left[\sin Q^{\rm T}\right]\>
\ex{i\> {\phi_+ +\phi_-\over2}}$, which leads to the Bogomol'nyi-like
bound
\begin{equation} 
M^{(-)}\geq {k\over \pi}\> \sqrt{|\lambda|}\;\; \bigl| \sin Q^{\rm
T} \bigr|\>.
\lab{BPS}
\end{equation}
For each value of $Q^{\rm T}$, this bound is saturated by the solutions to
the first order equation
\begin{equation}
\partial_x \psi =i \sqrt{-\lambda}\>\> {\rm sign\/}\left[\sin Q^{\rm
T}\right]\> \ex{i\>
{\phi_+ +\phi_-\over2}}\> (1- \psi
\psi^\ast)\>,
\lab{BacklundNeg}
\end{equation}
which yields the one-soliton solution~\rf{LundSol}. Its mass is given
by~\rf{MassForm}, which is clearly fixed by the charge $Q^{\rm
T}=(\phi_+-\phi_-)/2$. 

The duality relation~\rf{TdualSol} makes natural to ask whether there is an
analogue of the bound~\rf{BPS} in the $\lambda>0$ phase. We show below that
the answer to this question is affirmative, and that the one-soliton
solutions~\rf{ElectricSol} saturate another bound
characterised by their Noether charge, which provides a novel interpretation
for them as two-dimensional examples of Coleman's Q-balls~\cite{COLEMAN}.
This bound can be deduced as follows.

We start
with the energy density corresponding to a generic field configuration 
$\psi=\psi(t,x)$
\begin{equation}
{\cal H}_{\rm CSG} = {k\over 4\pi }\left({\partial_t
\psi\>\partial_t \psi^\ast+
\partial_x \psi\>\partial_x \psi^\ast
\over 1- \psi \psi^\ast} + \lambda\>
\psi \psi^\ast\right)\>.
\lab{CSGHamPlus}
\end{equation}
Using $x_\pm=t\pm x$ and $\partial_\pm = \partial/\partial x_\pm$,
it  can be witten as
\begin{eqnarray}
&&\hskip-0.5cm 
{\cal H}_{\rm CSG}= {k\over
8\pi}\biggl(\>{\bigl|2\>\partial_+
\psi\> -\ex{i\Gamma} \sqrt{\lambda}\> \psi\sqrt{(1- \psi
\psi^\ast)}\bigr|^2 + \bigl|2\>\partial_-
\psi\> +\ex{-i\Gamma} \sqrt{\lambda}\> \psi\sqrt{(1- \psi
\psi^\ast)}\bigr|^2
\over 1- \psi \psi^\ast}\nn[10pt]
&&\qquad\qquad\quad+ \>4 \sqrt{\lambda}\>{\rm
\> Re\/}\Bigl(\ex{i\Gamma}\> {\psi\partial_+
\psi^\ast - \psi^\ast\partial_-
\psi \over \sqrt{1- \psi \psi^\ast}}\Bigr)\biggr)\>,
\lab{CSGPlusB}
\end{eqnarray}
which leads to
\begin{equation}
E^{(+)}=\int_{-\infty}^{+\infty} dx\> {\cal H}_{\rm CSG} \geq
{k\over 2\pi}\>\sqrt{\lambda} \int_{-\infty}^{+\infty} dx\>\>{\rm
Re\/}\Bigl(\ex{i\Gamma}\> {\psi\partial_+
\psi^\ast - \psi^\ast\partial_-
\psi \over \sqrt{1- \psi \psi^\ast}}\Bigr)\>,
\lab{BPSPlusA}
\end{equation}
for an arbitrary function $\Gamma=\Gamma(t,x)$. Consider the
choice where $\Gamma$ is a solution to the
equations
\begin{equation}
\partial_\pm \Gamma = \pm \>{i\over2}\> {\psi
\partial_\pm \psi^\ast -\psi^\ast \partial_\pm \psi\over
1-\psi\psi^\ast}=\pm {2\pi\over k} J_\pm^{\rm N}\>,
\lab{GammaDef}
\end{equation}
which are integrable because the Noether current~\rf{NoetherCSG}
is conserved.
This way, and taking the boundary conditions~\rf{ElecBC} into account,
eq.~\rf{BPSPlusA} becomes 
\begin{eqnarray}
&&E^{(+)}\geq
-{k\over 2\pi}\>\sqrt{\lambda} \int_{-\infty}^{+\infty} dx\>\>
\partial_x {\rm
Re\/}\bigl(\ex{i\Gamma} \sqrt{1-\psi\psi^\ast}\bigr) \nn [2pt]
&& \qquad\quad =-
{k\over 2\pi}\>\sqrt{\lambda} \>\Bigl(\cos \Gamma(t,+\infty)-
\cos \Gamma(t,-\infty) \Bigr)\>.
\lab{BPSPlusB}
\end{eqnarray}
Notice that $\partial_x\Gamma=(2\pi/k) J_0^{\rm N}$, which means
\begin{equation}
\Gamma(t,+\infty)- \Gamma(t,-\infty) = {2\pi\over k}
\> \int_{-\infty}^{+\infty} dx\> J_0^{\rm N}= {2\pi\over k}\>  Q^{\rm
N}\>;
\lab{GammaCharge}
\end{equation}
otherwise the value of $\Gamma(t,\pm\infty)$ is
arbitrary. Therefore,
$E^{(+)}$ is actually bounded below by the maximal value of the
right-hand-side of~\rf{BPSPlusB}, which is attained for
$\Gamma(t,\pm\infty)= \pm \pi Q^{\rm N}/k -\pi \>{\rm
sign}\bigl[\sin(\pi Q^{\rm N}/k)\bigr]/2$.
This finally leads to the Bogomol'nyi-like bound
we were looking for
\begin{equation}
E^{(+)}\geq
{k\over \pi}\>\sqrt{\lambda}\>  \Bigl|\sin\left({\pi Q^{\rm N}\over
k}\right)\Bigr|\>.
\lab{BPSPlusC}
\end{equation}
For each value of the Noether charge $Q^{\rm N}$, the
bound~\rf{BPSPlusC} is saturated by the solutions to the
equations
\begin{equation}
2\>\partial_\pm
\psi= \pm \ex{\pm i\Gamma} \sqrt{\lambda}\> \psi\sqrt{(1- \psi
\psi^\ast)}\>,
\lab{BacklundPos}
\end{equation}
where $\Gamma$ is the solution to~\rf{GammaDef} with the boundary values
specified just before~\rf{BPSPlusC}.
This yields the one-solitons~\rf{ElectricSol}
originally constructed by Getmanov, whose mass is given
by~\rf{ChargeMass}, which is indeed fixed by the Noether charge. 

The Bogomol'nyi-like bounds deduced
in the previous paragraphs provide a novel alternative characterisation of the
previously known one-solitons solutions~\rf{LundSol} and~\rf{ElectricSol}
that makes their classical stability explicit.
The reason is that the solutions whose energy saturates a
bound like~\rf{BPS} or~\rf{BPSPlusC} are, in general, below threshold
for decay into a multi-particle state of the same charge $Q^{\rm T}$ or
$Q^{\rm N}$, respectively. In principle, they could still be at threshold for decay in other
states that saturate the same bound. However, say for~\rf{BPS}
and~\rf{MassForm}, this could only occur if
\begin{equation}
Q^{\rm T}= Q^{\rm T}_1+ Q^{\rm T}_2 \qquad\text{and} \qquad M\left(Q^{\rm
T}_1+ Q^{\rm T}_2\right)= M\left(Q^{\rm T}_1\right) + M\left(Q^{\rm
T}_2\right)\>
\end{equation}
but, taking into account the
identification $Q^{\rm T}\sim Q^{\rm T} + \pi{\fl Z}$ deduced just
after~\rf{TdualSol}, it is straightforward to check that the only
solution to these conditions is the trivial one where either
$Q^{\rm T}_1$ or
$Q^{\rm T}_2$ vanishes. Therefore, the solutions that saturate the 
bound in each phase, which in this case are the one-soliton
solutions~\rf{ElectricSol} and~\rf{LundSol}, are indeed classically
stable.

Finally, let us point out that the first order partial
differential equations~\rf{BacklundNeg} and~\rf{BacklundPos} satisfied
by the configurations that saturate the Bogomol'nyi-like
bounds~\rf{BPS} and~\rf{BPSPlusC} coincide with the B\"acklund
equations used in~\cite{CSGDUALITY} to construct the one-soliton solutions.

\sect{Conclusions}
\label{Conclu}

~\indent
We have studied T-duality in a
family of massive integrable field
theories associated with cosets of the form $G/(H\times U(1))$ whose 
classical equations of motion are particular examples of the non-abelian
affine Toda equations of Leznov and Saveliev~\cite{NAAT}. Among others, the
family includes the complex sine-Gordon model~\cite{LUND,POHL,BAKAS}, the
Homogeneous and Symmetric Space sine-Gordon models constructed
in~\cite{HSGClass,HSGQuan1,HSGQuan2,SSSG}, and the models associated with
cosets of  the form $SL(2)\times U(1)^n/U(1)$ considered in~\cite{GOMES1}. 
For each coset, the different theories are defined in terms
of an asymmetric gauged WZW action~\cite{GWZW,ASYM} perturbed by a potential.
Following~\cite{HSGClass}, the different gauged left and right actions of
$H\times U(1)$ on $G$ are labelled by an automorphism $\tau$ of
$H\times U(1)$.

Our main result is that all these theories exhibit off-shell abelian
T-duality symmetries. For each $U(1)$ generator in $H\times U(1)$, there is a
duality transformation given by eqs.~\rf{U1Dual}
and~\rf{U1DualAct} that relates two different theories associated with
the same coset. These theories admit an equivalent description in
terms of massive non-linear sigma models. Then, the form of the duality
transformations is summarised by eq.~\rf{TDualMass}, which exhibits that the
potential remains invariant, and that the transformation is actually a
consequence of the duality symmetries of the unperturbed theory.

In some cases, the two dual unperturbed theories coincide, their
Lagrangian actions are related by a change of field variables, and this change
modifies the potential. Then, the duality transformation has a different
interpretation. It relates two perturbations of the same self-dual
conformal field theory by different potentials. The form of this kind of
duality transformations is summarised by eq.~\rf{TPoint}. 

Explicit examples
of this type are provided by the HSG models, which
are associated to cosets of the form
$G/U(1)^r$, where $G$ is a compact simple Lie group. In this case, our
results show that they exhibit a duality transformation of the
form~\rf{TPoint} for each Weyl transformation of $G$ given by~\rf{DualAct},
where the potential transforms according to~\rf{PotTransf}. 
These transformations relate different phases of the
models, which we have also characterised by studying the
manifold of vacuum field configurations. They are in one-to-one
relation with the elements of the Weyl group of $G$.
We also
conjecture that these duality symmetries survive in the quantum theory, which
would be interesting to check against the available non-perturbative
results for the HSG models; in particular, those recently obtained by means
of the thermodynamic Bethe ansatz in~\cite{PATRICK}.

On-shell, the T-duality transformations provide a map between the solutions to
the classical equations of motion of the dual models. 
Since they exhibit, at least, a global
$U(1)$ symmetry leading to a conserved Noether current, their equations of
motion have soliton solutions that carry a finite value of the corresponding
$U(1)$ charge. Restricted to them, the duality map
turns Noether soliton charges into topological ones, which is reminiscent
of the string theory case, where T-duality changes trivial boundary conditions
into non-trivial ones.

We have studied in detail this transformation in the particular case of the
CSG model, which corresponds to the coset $SU(2)/U(1)$. It has two
different phases corresponding to the two signs of it unique coupling
constant, and solitons are known to be not of topological nature in one of
them. Then, the duality transformation suggests a topological interpretation
for them in the dual phase. Although  the conservation of the relevant
topological charge is not truly of topological nature, this interpretation
leads to a previously unreported Bogomol'nyi-like bound for the energy that
is saturated by these solitons. In the other, non-topological, phase, we also
show that the already known one-soliton solutions saturate another
Bogomol'nyi-like bound, which provides a novel interpretation for them as
Q-balls~\cite{COLEMAN}, and makes their classical stability explicit. 

The HSG models can be seen as generalisations of the CSG model. In fact, the
one-solitons solutions of the $G/U(1)^r$ HSG model have been constructed
in~\cite{HSGSemi} by embedding the non-topological $SU(2)$ CSG solitons
into the regular
$SU(2)$ subgroups of $G$. Taking into account the description of the phases of
the HSG models presented in section~\ref{HSG}, the construction
of~\cite{HSGSemi} was performed in a phase where the vacuum configuration is
not degenerate, similar to the non-topological phase of the CSG
model. Therefore, the resulting solitons are not topological, and
their classical stability is not clearly established. In spite of this, their
properties determine the quantum HSG theories constructed in~\cite{HSGQuan2},
and it would be extremely interesting to extend the results achieved in the
CSG model to this case. In particular, to investigate the possible
existence of Bogomol'nyi-like bounds saturated by the HSG solitons which
could clarify their stability.

\vspace{0.5 truecm}

\noindent
\parbox{\linewidth}{
\centerline{\large\bf Acknowledgments} 

\vspace{0.2truecm}

\hspace{0.5truecm}
I would like to thank Patrick Dorey, Frank Gomes, Tim Hollowood, 
Joaqu\'\i n S\'anchez Guill\'en, and Alfonso V\'azquez Ramallo for
helpful discussions during the course of this project. This research was
partly supported by MCyT (Spain) and FEDER (BFM2002-03881 and FPA2002-01161),
Incentivos from Xunta de Galicia, the NATO grant\break PST.CLG.980424, and the
EC network ``EUCLID" under contract HPRN-CT-2002-00325.}




\begin{thebibliography}{99}
\bibitem{TDUALITY} A.~Giveon, M.~Porrati and E.~Rabinovici, `Target space
duality in string theory',
\PR{244}{1994}{77}, \hepth{9401139};\\[5pt]
%
E.~Alvarez, L.~Alvarez-Gaum\'e and Y.~Lozano, `An introduction to T duality
in string theory', {\sl Nucl. Phys. B (Proc. Suppl.)} {\bf 41} (1995)~1,
\hepth{9410237};\\[5pt]
%
O.~Alvarez, `Target space duality. I: General theory', \NPB{584}{2000}{659},
\hepth{0003177}; `Target space duality. II: Applications', {\sl ibid.} {\bf B
584} (2000) 682, \hepth{0003178}.
\bibitem{HSGClass} C.R.~Fern\'andez-Pousa, M.V.~Gallas, T.J.~Hollowood
and J.L.~Miramontes, `The symmetric space and homogeneous sine-Gordon
theories', \NPB{484}{1997}{609}, \hepth{9606032}.
\bibitem{HSGQuan1} C.R.~Fern\'andez-Pousa, M.V.~Gallas, T.J.~Hollowood and
J.L.~Miramontes, `Solitonic integrable perturbations of parafermionic
theories', \NPB{499}{1997}{673}, \hepth{9701109}.
\bibitem{HSGQuan2} J.L.~Miramontes and C.R.~Fern\'andez-Pousa,
`Integrable quantum field theories with unstable
particles', \PLB{472}{2000}{392-401}, \hepth{9910218}; \\[5pt] 
%
C.~Korff,  `Colours
associated to non simply-laced Lie algebras
and exact S-matrices', \PLB{501}{2001}{289-296}, \hepth{0010287}.
\bibitem{LUND} F.~Lund and T.~Regge, `Unified approach to
strings and vortices with soliton solutions', \PRD{14}{1976}{1524-1535};
\\[5pt]
%
F.~Lund, `Example of a relativistic, completely
integrable, hamiltonian system', \PRL{38}{1977}{1175}.
\bibitem{POHL} K.~Pohlmeyer, `Integrable hamiltonian
systems and interactions through quadratic constraints', \CMP{46}{1976}{207}.
\bibitem{BAKAS} I.~Bakas, `Conservation laws and
geometry of perturbed coset models', \IJMPA{9}{1994}{3443-3472},
\hepth{9310122}.
\bibitem{PARK} Q-H.~Park, \PLB{328}{1994}{329-336}, `Deformed
coset models from gauged WZW actions', \hepth{9402038}.
\bibitem{SSSG}
O.A.~Castro Alvaredo and J.L.~Miramontes, \NPB{581}{2000}{643},
`Massive Symmetric Space sine-Gordon Soliton Theories and Perturbed Conformal
Field Theory', \hepth{0002219}.
\bibitem{GOMES1} J.F.~Gomes, E.P.~Gueuvoghlanian, G.M.~Sotkov and
A.H.~Zimerman, `T-duality of axial and vector dyonic integrable
models', \AoP{289}{2001}{232-250}, \hepth{0007116}; `Soliton spectrum of
integrable models with local symmetries', \JHEP{0207}{~2002}{001},
\hepth{0205228};\\[5pt]
%
J.F.~Gomes, G.M.~Sotkov and A.H.~Zimerman, `Axial
vector duality in affine NA Toda models', in `Workshop on Integrable
Theories, Solitons and Duality', PrHEP-unesp2002/045,
\hepth{0212046}.
\bibitem{KIR} E. Kiritsis, `Duality in gauged WZW
models', \MPLA{6}{1991}{2871-2879};\\[5pt]
%
R.~Dijkgraaf, E.~Verlinde and
H.~Verlinde, `String propagation in a black hole geometry',
\NPB{371}{1992}{269}.
\bibitem{KIRplus} E. Kiritsis, `Exact duality symmetries
in CFT and string theory', \NPB{405}{1993}{109-142}.
\bibitem{ZAMO} A.B.~Zamolodchikov, `Integrable field theory from conformal
field theory', Advanced Studies in Pure Mathematics {\bf 19} (1989) 641-674.
\bibitem{KW} H.A.~Kramers and G.H.~Wannier, `Statistics of the
two-dimensional ferromagnet. Part 1', \PRv{60}{1941}{252}.
\bibitem{BUSCHER} T.H.~Buscher, `A symmetry of the string
background field equations', \PLB{194}{1987}{59-62}; `Path integral derivation
of quantum duality in nonlinear sigma-models',
\PLB{201}{1988}{466-472}.
\bibitem{AGcan} E.~Alvarez, L.~Alvarez-Gaum\'e and Y.~Lozano, `A Canonical
approach to duality transformations',
\PLB{336}{1994}{183}, \hepth{9406206}.
\bibitem{VENEZCan} A.~Giveon, E.~Rabinovici and G.~Veneziano, `Duality In
String Background Space', \NPB{322}{1989}{167}.
\bibitem{GOMES2} J.F.~Gomes, E.P.~Gueuvoghlanian, G.M.~Sotkov and
A.H.~Zimerman, `Electrically charged topological solitons',
\NPB{606}{2001}{441-482}, \hepth{0007169}; `Dyonic sigma models',
\NPB{598}{2001}{615-644}, \hepth{0011187}; `Soliton spectrum of
integrable models with local symmetries', \JHEP{0207}{2002}{001},
\hepth{0205228};\\[5pt]
%
I.~Cabrera-Carnero, J.F.~Gomes, G.M.~Sotkov and
A.H.~Zimerman, `Multicharged dyonic integrable models',
\NPB{634}{2002}{433-482}, \hepth{0201047}; `Vertex operators and
soliton solutions of affine Toda model with $U(2)$ symmetry',
\JPA{37}{2004}{6375-6390}, \hepth{0403042}; \\[5pt]
%
J.F.~Gomes, G.M.~Sotkov and A.H.~Zimerman, `T-duality in 2D integrable
models', \JPA{37}{2004}{4629-4640}, \hepth{0402091}; 
`Solitons with isospin',
\hepthP{0405182}.
\bibitem{PARKSHIN} Q-H.~Park and H.J.~Shin, `Deformed minimal models and
generalized Toda theory',
\PLB{347}{1995}{73}, \hepth{9408167}; `Classical
matrix sine-Gordon theory', \NPB{458}{1996}{327-354},
\hepth{9505017}.
\bibitem{TLWZW} L.A.~Ferreira, J.L.~Miramontes and J.~S\'anchez Guill\'en,
\NPB{449}{1995}{631}, `Solitons, Tau-functions and Hamiltonian Reduction for
Non-Abelian Conformal Affine Toda Theories',
\hepth{9412127};\\[5pt]
%
L.A.~Ferreira, J.L.~Gervais, J.~S\'anchez Guill\'en and
M.V.~Saveliev, `Affine Toda Systems Coupled to Matter Fields',
\NPB{470}{1996}{236}, \hepth{9512105}.
%
\bibitem{SSSM} I.~Bakas, Q-H.~Park and H-J.~Shin,
`Lagrangian formulation of symmetric space sine-Gordon
models', \PLB{372}{1996}{45},
\hepth{9512030}.
%
\bibitem{MASSIVE} T.J.~Hollowood, J.L.~Miramontes and Q-H.~Park, `Massive
integrable soliton theories',
\NPB{445}{1995}{451}, \hepth{9412062}. 
\bibitem{NAAT} A.N.~Leznov and M.V.~Saveliev,
`Two-dimensional Exactly and Completely Integrable
Dynamical Systems', \CMP{89}{1983}{59-75}.
\bibitem{WZW} E.~Witten, `Non-abelian bosonization in
two dimensions', \CMP{92}{1984}{455-472}.
\bibitem{GWZW} D.~Karabali, Q.H.~Park, H.J.~Schnitzer and Z.~Yang,
`A GKO Construction based on a path integral formulation of gauged
Wess-Zumino-Witten actions', \PLB{216}{1989}{307};\\[5pt]
%
D.~Karabali and H.~J.~Schnitzer,
`BRST quantization of the gauged WZW action and coset conformal field
theories', \NPB{329}{1990}{649};\\[5pt]
%
K.~Gawedzki and A.~Kupianen, `$G/H$ conformal field theory
from gauged WZW model', \PLB{215}{1988}{119-123}; `Coset
construction from functional integrals', \NPB{320}{1989}{625-668};\\[5pt]
%
E.~Witten, `On Holomorphic Factorization of WZW Coset
Models', \CMP{144}{1992}{189-212}.
\bibitem{ASYM}
I.~Bars and K.~Sfetsos,
`Generalized duality and singular strings in higher dimensions',
\MPLA{7}{1992}{1091-1104}, [arXiv:hep-th/9110054];\\[5pt]
%
T.~Quella and V.~Schomerus, `Asymmetric cosets',
\JHEP{0302}{2003}{030}, [arXiv:hep-th/0212119];
`Asymmetrically Gauged WZNW Models',
{\em Fortsch.\ Phys.\/}  {\bf 51} (2003) 843-849.
\bibitem{PWTOP} G.~Felder, K.~Gawedzki and A.~Kupiainen,
`Spectra of Wess-Zumino-Witten models with arbitrary
simple groups', \CMP{117}{1988}{127-158};\\[5pt]
%
M.R.~Gaberdiel, `Abelian
duality in WZW models', \NPB{471}{1996}{217-232}, \hepth{9601016}.
\bibitem{ORLANDOPseudo} T.~Curtright and C.~Zachos, `Currents, charges, and
canonical structure of pseudodual chiral models',
\PRD{49}{1994}{5408-5421}, \hepth{9401006};\\[5pt]
%
O.~Alvarez, `Target space pseudoduality between dual symmetric
spaces', \NPB{582}{2000}{139}, \hepth{0004120}; `Pseudoduality in sigma
models', {\sl ibid.} {\bf B 638} (2002) 328, \hepth{0204011}.
\bibitem{HSGNp1} O.A.~Castro-Alvaredo, A.~Fring, C.~Korff and
J.L.~Miramontes, `Thermodynamic Bethe ansatz of the
homogeneous sine-Gordon models', \NPB{575}{2000}{535-560}, \hepth{9912196}.
\bibitem{HSGNp2} O.A.~Castro-Alvaredo, A.~Fring and C.~Korff,
`Form factors of the homogeneous sine-Gordon models',
\PLB{484}{2000}{167-176}, \hepth{0004089};\\[5pt]
%
O.A.~Castro-Alvaredo and A.~Fring, `Identifying the operator content,
the Homogeneous Sine-Gordon models', \NPB{604}{2001}{367}, \hepth{0008044};
`Renormalization group flow with unstable
particles', \PRD{63}{2001}{021701(R)},
\hepth{0008208}; `Decoupling the $SU(N)_2$-homogeneous sine-Gordon
Model', \PRD{64}{2001}{085007},
\hepth{0010262}.
\bibitem{PATRICK} P.~Dorey and J.L.~Miramontes,
`Mass scales and crossover phenomena in the Homogeneous Sine-Gordon Models', 
\hepthP{0405275}, to appear in {\em Nucl. Phys.~B}; `Aspects of the
homogeneous sine-Gordon models', in proceedings of the `Workshop on
Integrable Theories, Solitons and Duality', PrHEP-unesp2002/006,
[arXiv:hep-th/0211174].
\bibitem{CDF} O.A.~Castro-Alvaredo, J.~Dreissig and A.~Fring,
`Integrable scattering theories with unstable particles',
\EJPC{35}{2004}{393}, \hepth{0211168}.
\bibitem{OLIVE} P.~Goddard, J.~Nuyts and D.~Olive, `Gauge theories and
magnetic charge', \NPB{125}{1977}{1-28}.
\bibitem{HUMP} J.E.~Humphreys, `Introduction to Lie Algebras and
Representation Theory', New York-Berlin: Springer-Verlag, 1972.
\bibitem{CSGDUALITY} Q-H.~Park and H.J.~Shin,
`Duality in Complex sine-Gordon Theory', \PLB{359}{1995}{125},
\hepth{9506087}.
\bibitem{BT} P.~Bowcock and G.~Tzamtzis, {\em The complex sine-Gordon on 
a half line\/},\\ \hepthP{0203139}.
\bibitem{dVM} H.J.~de Vega and J.M.~Maillet,
`Renormalization character and quantum $S$-matrix for a
classically integrable theory', \PLB{101}{1981}{302-306}.
\bibitem{dVMSemi} H.J.~de Vega and J.M.~Maillet,
`Semiclassical quantization of the complex sine-Gordon
field theory', \PRD{28}{1983}{1441-1452}.
\bibitem{DH} N.~Dorey and T.J.~Hollowood, `Quantum
scattering of charged solitons in the complex sine-Gordon
model', \NPB{440}{1995}{215-233}, \hepth{9410140}.
\bibitem{PINT} V.A.~Fateev, `Integrable deformations
in ${\fl Z}_N$-symmetrical models of the conformal
quantum field theories', \IJMPA{6}{1991}{2109-21332};\\[5pt]
%
H.~de Vega and
V.A. Fateev, `Factorizable S Matrices For Perturbed W Invariant Theories',
\IJMPA{6}{1991}{3221}.
\bibitem{PARAF} A.B.~Zamolodchikov and V.A.~Fateev,
`Nonlocal (parafermion) currents in two-dimensional
conformal quantum field theory and self-dual critical
points in ${\fl Z}_N$-symmetric statistical systems', \JETP{62}{1985}{215}. 
\bibitem{CABRA} D.C.~Cabra and E.F.~Moreno, `Duality in
deformed coset fermionic models', \NPB{475}{1996}{522-534},
\hepth{9603138}.
\bibitem{DUALOLD} I.V.~Barashenkov and B.S.~Getmanov, `Multisoliton
solutions in the scheme of unified descrition of integrable relativistic
massive fieds. Non-degenerate $sl(2,{\fl C})$ case',
\CMP{112}{1987}{423-446}.
\bibitem{EUCLIDEAN} I.V.~Barashenkov and D.E.~Pelinovsky,
`Exact vortex solutions of the complex sine-Gordon
theory on the plane', \PLB{436}{1998}{117-124},
\hepth{9807045};\\[5pt]
%
I.V.~Barashenkov, V.S.~Shchesnovich and R.~Adams,
{\em Non-coaxial multivortices in the complex sine-Gordon theory on the
plane\/}, \nlinP{0202018}.
\bibitem{GETMANOV} B.S.~Getmanov, `New Lorentz-invariant
system with exact multisoliton solutions', \JETPL{25}{1977}{119}.
\bibitem{ParafCos} K.~Bardakci, M.~Crescimanno and E.~Ravinovici,
`Parafermions from coset models',
\NPB{344}{1990}{344-370}.
\bibitem{COLEMAN} S.~Coleman, `Q~balls', \NPB{262}{1985}{263}.
\bibitem{HSGSemi} C.R.~Fern\'andez-Pousa and J.L.~Miramontes,
`Semi-classical spectrum of the homogeneous sine-Gordon
theories', \NPB{518}{1998}{745-769}, \hepth{9706203}.
%




\end{thebibliography}
\end{document}